\documentclass[12pt,preprint]{aastex}
\usepackage{graphicx}
\begin{document}
\def\la{\mathrel{\hbox{\rlap{\hbox{\lower4pt\hbox{$\sim$}}}\hbox{$<$}}}}
\def\ga{\mathrel{\hbox{\rlap{\hbox{\lower4pt\hbox{$\sim$}}}\hbox{$>$}}}}
\def\lam{$\lambda$}
\def\kms{km~s$^{-1}$}
\def\vphot{$v_{phot}$}
\def\ang{~\AA}

\title {Direct Analysis of Spectra of the Peculiar Type~Ia Supernova 2000cx}

\author {David Branch\altaffilmark{1}, R.~C.~Thomas\altaffilmark{1,2},
   E.~Baron\altaffilmark{1}, Daniel~Kasen\altaffilmark{2},
   Kazuhito~Hatano\altaffilmark{3}, K.~Nomoto\altaffilmark{3},
   Alexei~V.~Filippenko\altaffilmark{4}, Weidong~Li\altaffilmark
   {4}, and Richard.~J.~Rudy\altaffilmark{5}}

\altaffiltext{1}{Department of Physics and Astronomy, University of
Oklahoma, Norman, Oklahoma 73019, USA; branch@nhn.ou.edu}

\altaffiltext{2}{Lawrence Berkeley National Laboratory, 1~Cyclotron
  Road, Berkeley, CA 94720--8158}

\altaffiltext{3}{Department of Astronomy and Research Center for the
  Early Universe, University of Tokyo, Bunkyo-ku, Tokyo, Japan}

\altaffiltext{4}{Department of Astronomy, 601~Campbell Hall,
University of California, Berkeley, CA 94720--3411}

\altaffiltext{5}{Space Science Applications Laboratory, The Aerospace
Corporation M2/266, P.~O.~Box 92957, Los Angeles, CA 90009}

\begin{abstract}

The Type~Ia SN~2000cx exhibited multiple peculiarities, including a
lopsided $B$--band light--curve peak that does not conform to current
methods for using shapes of light curves to standardize SN~Ia
luminosities.  We use the parameterized supernova synthetic--spectrum
code {\bf Synow} to study line identifications in the
photospheric--phase spectra of SN~2000cx.  Previous work established
the presence of Ca~II infrared--triplet features forming above
velocity $\sim$20,000 \kms, much higher than the photospheric velocity
of $\sim$10,000 \kms.  We find Ti~II features forming at the same high
velocity.  High--velocity line formation is partly responsible for the
photometric peculiarities of SN~2000cx: for example, $B$--band flux
blocking by Ti~II absorption features that decreases with time causes
the $B$ light curve to rise more rapidly and decline more slowly than
it otherwise would.

SN~2000cx contains an absorption feature near 4530\ang\ that may be
H$\beta$, forming at the same high velocity.  The lack of conspicuous
H$\alpha$ and P$\alpha$ signatures does not necessarily invalidate the
H$\beta$ identification if the high--velocity line formation is
confined to a clump that partly covers the photosphere and the
H$\alpha$ and P$\alpha$ source functions are elevated relative to that
of resonance scattering.  The H$\beta$ identification is tentative.
If it is correct, the high--velocity matter must have come from a
nondegenerate companion star.
     
\end{abstract}


\section{INTRODUCTION}

Supernova 2000cx in NGC~524 was an intriguing Type~Ia (SN~Ia) event.
[See Filippenko (1997) for an overview of supernovae.]  Li
et~al. (2001; hereafter L01) presented extensive optical photometric
and spectroscopic observations that revealed multiple peculiarities,
including the following:  (1) Around the time of maximum brightness
the $B$--band light curve rose rapidly but declined slowly, resulting
in a lopsided light--curve shape that does not conform to present
SN~Ia light--curve fitting techniques such as the multicolor
light--curve--shape (MLCS) method of Riess, Press, \& Kirshner (1996)
or the stretch method of Perlmutter et~al. (1997) and Goldhaber
et~al. (2001). (2) Around maximum brightness the $B - V$ color was
redder than that of a normal SN~Ia, but after about 15 days
postmaximum $B-V$ was bluer than normal. (3) In certain respects
(strong Fe~III, weak Si~II and S~II) the premaximum spectra resembled
those of the peculiar overluminous SN~1991T (Filippenko et~al. 1992a),
but the postmaximum spectral evolution differed from that of both
SN~1991T and normal SNe~Ia.  Fe~III lines persisted for an unusually
long time after maximum brightness while Fe~II lines were late to make
their appearance.  L01 discussed the implications of their
observations for the physical nature of SN~2000cx and tentatively
suggested that it may have been an overluminous event with an even
larger yield of $^{56}$Ni and a higher ejecta kinetic energy than
SN~1991T.

Rudy et~al. (2002; hereafter R02) obtained infrared spectra of
SN~2000cx, 5 and 6 days before maximum brightness, covering the range
0.8 to 2.5~$\mu$m.  Following Meikle et~al. (1996) and Wheeler
et~al. (1998), who attributed an absorption near 1.05~$\mu$m in the
normal SN~Ia 1994D to Mg~II \lam10296, R02 identified a 1.02~$\mu$m
absorption in SN~2000cx with the same transition, but forming at
unusually high velocity.  They inferred that freshly synthesized
magnesium extended to velocities higher than 20,000 \kms, and
discussed this finding as support for a delayed detonation explosion
model, rather than a deflagration model, for SN~2000cx.

Candia et~al. (2003; hereafter C03) presented additional optical and
infrared photometry and emphasized that SN~2000cx displayed some
characteristics of an {\sl underluminous} SN~Ia, including a rather
fast bolometric light curve.  With a recession velocity of only
$\sim$2200 \kms, NGC~524 is not in the Hubble flow, and various ways
of estimating its distance span a wide range, so C03 were not able to
decide whether SN~2000cx actually was overluminous or underluminous.

In Thomas et~al. (2003; hereafter T03) we concentrated on exploring
simple explanations for high--velocity ($\sim22,000$ \kms) absorption
features produced by the Ca~II infrared triplet.  Our favored solution
was a high--velocity non--spherical structure (hereafter a ``clump")
that only partly covered the photosphere.  We also noted that H$\beta$
might be present in absorption at the same high velocity.

Some of the line identifications in the articles cited above were
tentative, and many of the spectral features were not identified at
all.  Here we use the parameterized supernova synthetic--spectrum code
{\bf Synow} to study line identifications in the photospheric--phase
spectra of SN~2000cx.  We also try to shed some light on the causes of
the peculiar $B$ light curve and $B-V$ color curve.

\section{AN OVERVIEW OF THE SPECTRA AND THE ANALYSIS}

The spectra we have studied are shown in Figure~1.  The optical
spectra, from L01, range from their earliest spectrum, obtained 3 days
before the time of maximum brightness in the $B$ band, to 32 days
after maximum, when the spectrum was beginning to make the transition
from the photospheric to the nebular phase.  The infrared spectrum is
the sum of the two spectra obtained by R02, 5 and 6 days before
maximum; hereafter we will refer to this as the day~$-5.5$ spectrum.
A few of the familiar SN~Ia absorption features in the optical,
produced by Si~II, S~II, and Ca~II, are labelled.  In Figure~1 the
logarithm of the wavelength is plotted, to allow a fair comparison of
the widths of the spectral features at different wavelengths.  In
subsequent figures the more traditional practice of plotting
wavelength is followed.  As Figure~1 shows, the spectra did not change
very much between days $-3$ and 7, although there are some significant
differences that we will examine below.  The changes between days~7
and 15, and especially between days 15 and 32, are much more obvious.

In a recent article (Branch et~al. 2003; hereafter B03) we used {\bf
Synow} to carry out a direct analysis of early spectra of the normal
SN~Ia 1998aq.  Instead of repeating the discussion of the code and the
input parameters, we refer the reader to B03.  In \S3 we present a
detailed discussion of the day~2 spectrum of SN~2000cx.  In \S4 we
more briefly discuss the day~7, 15, and 32 spectra.  In \S5 we examine
the premaximum spectra, beginning with the day~$-3$ spectrum.  When
analyzing the optical spectra we restrict our attention to wavelengths
shorter than 7000\ang, since the longer wavelength region containing
the Ca~II infrared triplet was discussed by T03.  The day~$-5.5$
infrared spectrum also is considered in \S5.  The results are
discussed in \S6.

\section{THE DAY 2 SPECTRUM}

The day~2 spectrum of SN~2000cx is shown in Figure~2 with its absorption
features labelled by wavelength.  The same spectrum is compared with a
day~2 spectrum of SN~1998aq in Figure~3.  Ions used by B03 to fit the
absorption features in SN~1998aq are indicated.  By the standards of
SNe~Ia near maximum light, the differences between these two spectra
are large.  In SN~2000cx, the absorptions that the two events have in
common are more blueshifted, and the prominent Si~II and S~II
absorptions are weaker.  SN~2000cx has distinct absorptions at 3900
and 4530\ang\ that are missing or weaker in SN~1998aq, while SN~1998aq
has an absorption at 4420\ang, attributed by B03 to Si~III, that is
missing or weaker in SN~2000cx.  Clearly the main reason that the
$B-V$ color of SN~2000cx was redder than that of SN~19998aq at this
epoch is that from 3900 to 4300\ang\ the spectrum of SN~2000cx is
strongly depressed compared to that of SN~1998aq.  The weaker S~II
absorptions of SN~2000cx also contribute to making its $B-V$
redder. [It should be mentioned that although SN~1998aq had normal
SN~Ia spectral features, its $B-V$ color was somewhat bluer than
average (Boffi \& Riess 2003; A.~G.~Riess et~al., in preparation)].

We used {\bf Synow} to calculate numerous synthetic spectra for
comparison with the day~2 spectrum, considering all ions that can be
regarded as plausible according to the supernova spectrum atlas of
Hatano et~al. (1999b).  The fitting process was unusually lengthy
because it proved necessary to consider line formation at both
photospheric velocities (PV, $\sim$10,000 \kms) and at high velocity
(HV, $\sim$20,000 \kms).  Rather than using excitation temperatures as
fitting parameters we used the ion--specific default values (see B03)
for PV line formation and a nominal value of 5000~K for HV line
formation.  We began with $v_e =1000$ \kms\ as the default
optical--depth e--folding velocity, but (as reported in the tables) in
some instances we used larger values of $v_e$ to improve the fits.

Figure~4 shows a comparison of the day~2 spectrum with a synthetic
spectrum that has \vphot=11,000 \kms, $T_{bb}=9000$~K, and contains
lines of nine ions: Mg~II, Si~II, S~II, Fe~III, Co~II, and Ni~II have
only PV components; Ca~II has both PV and HV components; Ti~II and
Fe~II have only HV components.  The ion--specific input parameters ---
the wavelength and maximum optical depth of the reference line, the
minimum, maximum, and optical--depth e--folding velocities, and the
excitation temperature --- are listed in Table~1.  Except for Ca~II
the PV ions are mildly detached from the photosphere, at 13,000 \kms.
(Detached refers to non--zero line optical depths beginning not at the
photospheric velocity but at a higher detachment velocity.  By mildly
detached we mean that the difference between the two velocities
is rather small.)  When these ions are undetached at 11,000 \kms\
their absorptions are not blueshifted enough.  When $v_{phot}$ is
increased from 11,000 to 13,000 \kms, so that these ions become
undetached at 13,000 \kms, the red edge of the Ca~II H\&K absorption
is too blueshifted and most of the features are slightly too broad.
It is not clear, however, that the small difference between the
adopted photospheric velocity of 11,000 \kms\ and a detachment
velocity of 13,000 \kms\ is physically significant; it may simply
reflect the limitations of {\bf Synow} (e.g., the sharp photosphere
assumption).

The overall fit in Figure~4 is fairly good, but there are several
obvious discrepancies.  As usual with SNe~Ia, the observed spectrum is
more strongly peaked in the middle of the optical spectrum than a
blackbody spectrum.  We use a blackbody continuum temperature $T_{bb}$
that works well for the middle of the spectrum because that is where
the most interesting line identification issues (HV Ti~II, H$\beta$;
see below) turn out to be.  Therefore the red end of the synthetic
spectrum, beginning near 5300\ang, is too high.  Obtaining a good fit
to the shape of the overall energy distribution is a task for more
detailed spectrum calculations, such as those of Lentz et~al. (2001).
The synthetic spectrum has a peak at 3950\ang\ that is too high, and
it does not fit well at wavelengths shorter than 3600\ang; the fit
could be improved by introducing more iron--peak ions, but with more
adjustable parameters.  The most interesting discrepancy is the lack
of a synthetic counterpart to the 4530\ang\ absorption.  We will
return to this below.

Figure~5 is like Figure~4 but with only the PV components in the
synthetic spectrum.  The long--wavelength half of the synthetic
spectrum is dominated by PV features of Si~II, S~II, and Fe~III, but
at shorter wavelengths the PV--only fit is incomplete.  To achieve the
fit of Figure~4, PV Ni~II is needed for the 3900\ang\ absorption (one of
the absorptions that is not seen distinctly in SN~1998aq; Figure~3) and
PV Co~II is needed for the 4000\ang\ absorption, but as Figure~5 shows,
this part of the spectrum is not well fit with PV components alone.
In Figure~5 there also is a serious lack of synthetic absorption to the
right of what appears to be a sharp emission peak in the observed
spectrum at 4060\ang.  L01 called attention to this apparent emission
feature and noted that it had not previously been observed in SNe~Ia.

Figure~6 is like Figure~4 but with only the HV components in the
synthetic spectrum.  The HV Fe~II lines play a relatively minor role
and their identification is not definite, but remarkably, HV Ti~II
lines are important.  These are the same Ti~II lines that, forming at
lower velocities, produce the blue trough in peculiar underluminous
SN~1991bg--like SNe~Ia (Filippenko et~al. 1992b), in SNe Ib (Branch
et~al. 2002), and in SNe~Ic (Millard et~al. 1999).  The difference
between the signatures of PV Ti~II, as seen in SN~1991bg--like SNe~Ia,
and HV Ti~II, as seen in SN~2000cx, is shown in Figure~7.  Note that
in both cases Ti~II features appear near the middle of the $B$ band.
The HV Ti~II lines of SN~2000cx produce both a sudden depression of
the spectrum beginning at wavelengths shorter than 4290\ang\ and a
flux peak at 4060\ang.  The previously unidentified 4060\ang\ peak
turns out to be part of the signature of blended HV Ti~II
resonance--scattering features.  (The identification of HV Ti~II in
SN~2000cx is a good example of the utility of {\bf Synow}.  It is
unlikely that HV Ti~II could have been recognized by looking at line
lists or by performing detailed spectrum calculations for existing
hydrodynamical models.)  The HV Ti~II features of SN~2000cx are partly
responsible for the depression of the spectrum from 3900 to 4300\ang,
therefore they alter the $B$ magnitude and the $B-V$ color relative to
normal SNe~Ia, which do not have significant Ti~II features.

The synthetic spectrum of Figure~4 does not account for the 4530\ang\
absorption.  We have not been able to find a plausible PV ion to
account for this absorption.  A transition blueshifted by about 13,000
\kms\ would need to have a rest wavelength near 4720~\AA.  The best
fit we could achieve with PV matter, without seriously degrading the
fit elsewhere, would be with undetached He~II \lam4686, but this
identification is unlikely because even with nonthermal excitation by
radioactivity the appearance of He~II lines is unexpected, and we
would need to invoke He~II forming at a lower velocity than all other
PV ions except Ca~II.  When He~II is detached to 13,000 \kms, like the
other PV ions, its absorption is too blue to fit the 4530\ang\
absorption.  The strongest optical line of C~III, \lam4647, is too
blue even when it is undetached at 11,000 \kms.  C~II \lam4745
detached at 15,000 \kms\ could fit the 4530\ang\ absorption, but then
C~II \lam6580 would produce a strong feature near 6250\ang\ that is
not observed.

If the 4530\ang\ absorption is not produced by PV matter then perhaps
it is produced by HV matter. A transition blueshifted by about 22,000
\kms\ would need to have a rest wavelength near 4860~\AA.  The only
plausible HV candidate that we can suggest is H$\beta$, at \lam4861.
Figure~8 shows the effect of including hydrogen lines detached at
22,000 \kms, with an optical depth of the reference line, H$\alpha$,
of 5.  With these parameters H$\beta$ accounts for the 4530\ang\
absorption, while H$\alpha$ falls within the 6110\ang\ absorption.  At
first glance the excessive depth of the synthetic H$\alpha$ absorption
appears to contradict the H$\beta$ identification, but as discussed in
T03, this is not necessarily the case.  If HV hydrogen is in a clump
that covers only a fraction of the photosphere, as inferred by T03 for
the HV Ca~II absorption, rather than covering all of the photosphere
as assumed in a 1-D {\bf Synow} calculation, then the fractional depth
of the H$\alpha$ absorption is limited by the photospheric covering
fraction.  This can be seen in Figure~13 of T03, where H$\beta$ and
H$\alpha$ are simultaneously fit reasonably well by a model in which
hydrogen is confined to a HV clump with a low photospheric covering
fraction.

Since we have not found a plausible alternative to H$\beta$, forming
at {\sl any} velocity between 13,000 and 22,000 \kms, in the remainder
of this article we will use H$\beta$ to account for the 4530\ang\
absorption.  We will return to the issue of H$\alpha$, and consider
the Paschen lines, in \S5.1 and \S5.2.

\section{LATER SPECTRA}

In this section we more briefly discuss the day~7, day~15, and day~32
spectra.  At each epoch we compare the spectrum of SN~2000cx with (1)
a spectrum of SN~1998aq; (2) an earlier--epoch spectrum of SN~2000cx;
and (3) a synthetic spectrum.

\subsection{The Day~7 Spectrum}

The day~7 spectrum of SN~2000cx is compared with a day~7 spectrum of
SN~1998aq in Figure~9.  The differences remain substantial and generally
similar to those of day~2.  Again the suppression of the blue part of
the spectrum and the weaker S~II lines cause the $B-V$ color of
SN~2000cx to be redder than that of SN~1998aq.

The day~7 and day~2 spectra of SN~2000cx are compared in Figure~10.  The
wavelengths of the absorptions are very similar.  The most significant
difference is that the day~7 spectrum has developed more prominent
peaks near 4450 and 5170\ang, with the latter apparently partly
filling in the 5270\ang\ absorption.

Figure~11 shows a comparison of the day~7 spectrum with a synthetic
spectrum that has \vphot=11,000 \kms, $T_{bb}=10,000$~K, and contains
lines of the same ten ions as in Figure~8 (including H~I).  The
ion--specific input parameters are listed in Table~2.  Again all PV
ions except Ca~II are mildly detached at 13,000 \kms.  Comparing
Tables~1 and 2, we see that the reference--line optical depths have
changed only mildly.  The increase in the Fe~III optical depth and the
decrease in the S~II optical depth are responsible for getting
reasonable fits to the 4450 and 5170\ang\ peaks.  At day~7 the
synthetic spectrum in the blue still is a composite of PV and HV line
formation.

\subsection{The Day 15 Spectrum}

The day~15 spectrum of SN~2000cx is compared with a day~18 spectrum of
SN~1998aq in Figure~12.  Except at the red end, the spectra still are
very different.  In fact, at wavelengths shorter than about 4600\ang\
the spectra are practically opposite: SN~2000cx has absorptions where
SN~1998aq has peaks.  The spectra differ so much in the $B$ and $V$
bands that any close similarity of their $B-V$ colors would be
accidental.

The day~15 and day~7 spectra of SN~2000cx are compared in Figure~13.  At
day~15 the 6110\ang\ absorption has hardly changed, but a broad
absorption has developed at 5670\ang\ and the S~II absorptions no
longer can be seen.  The 4530\ang\ absorption is still present, now to
the left of an emission peak that has emerged at 4580\ang.

Figure~14 shows a comparison of the day~15 spectrum with a synthetic
spectrum that has \vphot=10,000 \kms, $T_{bb}=12,000$~K, and contains
lines of seven ions: Na~I, Si~II, Fe~II, Fe~III, and Co~II have PV
components; H~I and Ti~II are at HV.  At this epoch Mg~II, S~II, and
Ni~II lines are no longer being used.  The ion--specific input
parameters are listed in Table~3.  The broad 5670\ang\ absorption is
well fit by the Na~I D lines (\lam\lam5890,5896) with a high value of
$v_e=7$.  The spectrum at wavelengths shorter than 5000\ang\ also is
well fit.  This has been accomplished by retaining the Fe~III lines
and by introducing Fe~II lines having a significant optical depth
throughout the whole velocity interval of 11,000 to 23,000 \kms\ (by
using a high e--folding velocity of $v_e=20,000$ \kms).  The
development of the Fe~II features causes the emergence of the
4580\ang\ peak.  The large differences between the spectra of
SNe~2000cx and 1998aq at this epoch are due to the persistent Fe~III
features and to Fe~II features forming throughout the wide velocity
interval.

\subsection{The Day 32 Spectrum}

The day~32 spectrum of SN~2000cx is compared with a day~32 spectrum of
SN~1998aq in Figure~15.  At this epoch the spectra generally contain the
same features, but those in the blue remain more blueshifted in
SN~2000cx.  Both L01 and C03 emphasized that SN~2000cx violates the
``Lira--Phillips'' law (Lira 1995; Phillips et~al. 1999) --- that the
extinction--corrected $B-V$ color of SNe~Ia evolves in an uniform way
from day~30 to day~90.  Throughout this time interval SN~2000cx is
anomalously blue.  It is clear from Figure~15 that in addition to
whatever peculiarity there may be in the underlying continuum (or
pseudo--continuum), the higher blueshifts of the features in SN~2000cx
play a role in altering the $B-V$ color at day~32.

The day~32 and day~15 spectra of SN~2000cx are compared in Figure~16.
At day~32 the overall energy distribution is much redder.  The
6110\ang\ absorption is still present but flanked by absorptions
at 6000 and 6280\ang\ that are commonly attributed to Fe~II.  L01
pointed out that these Fe~II features were unusually late to develop
in SN~2000cx.  At wavelengths shorter than about 5400\ang\ the
spectrum has changed dramatically: broad absorptions appear near
5100\ang\ and 4400\ang, where emission peaks appeared at day~15.  The
4530\ang\ absorption no longer is seen, although its presence on the
steep blue side of the 4600\ang\ peak is difficult to exclude.

Figure~17 shows a comparison of the day~32 spectrum with a synthetic
spectrum that has \vphot=11,000 \kms, $T_{bb}=8000$~K, and contains
lines of Na~I, Si~II, Ca~II, and Fe~II at PV, and only Ca~II at HV.
H~I and Fe~III lines are no longer used.  The ion--specific input
parameters are listed in Table~4.  Apart from the three features
labelled in Figure~17, all of the synthetic features are due to Fe~II.
As we did for the late photospheric spectra of SN~1991T (Fisher et
al. 1999) and of the SN~1991T--like event SN~1997br (Hatano
et~al. 2002), we have introduced a discontinuity in the Fe~II optical
depth.  It decreases exponentially from 100 at $v_{phot} = 11,000$
\kms\ to 37 at $v=13,000$ \kms; there it abruptly falls to 2 and then
decreases exponentially again.  The broad 5100 and 4400\ang\
absorptions are formed in the high optical--depth region, which we
interpret as the outer part of an iron--rich core.  (In SNe~1991T and
1997br we placed the discontinuity at 10,000 and 12,500 \kms,
respectively.)  The differences between SNe~2000cx and 1998aq in
Figure~17 are mainly caused by Fe~II line formation extending to higher
velocities, and to deep HV Ca~II, in SN~2000cx.

\section{PREMAXIMUM SPECTRA}

In this section we turn our attention to the premaximum spectra: the
optical day~$-3$ spectrum and the infrared day~$-5.5$ spectrum.

\subsection{The Day $-3$ Spectrum}

The day~$-3$ spectrum of SN~2000cx is compared with a day~$-3$
spectrum of SN~1998aq in Figure~18.  The difference in the blue is
qualitatively like at day~2, but even more extreme.

The day~$-3$ and day~2 spectra of SN~2000cx are compared in Figure~19.
The day~$-3$ spectrum has a distinct absorption feature at 4370\ang\
and a weaker one at 5510\ang\, both of which are weak or absent in the
day~2 spectrum.  Otherwise, the same spectral features are present but
at different strengths: at day~$-3$ the S~II features are much weaker
and the spectrum in the blue is even more suppressed than at day~2.

Figure~20 shows a comparison of the day~$-3$ spectrum with a synthetic
spectrum that has \vphot=12,000 \kms, $T_{bb}=9000$~K, and contains
lines of nine ions: the same as in Figure~7 for the day~2 spectrum
except that PV Co~II has been omitted and PV Si~III has been added.
The ion--specific input parameters are listed in Table~5.  The
4370\ang\ absorption is accounted for by Si~III \lam4550.  L01
suggested that the 5510\ang\ absorption could be Si~III \lam5740.  We
believe this to be correct, because although the synthetic spectrum of
Figure~20 does not produce the feature, if we make \lam4550 somewhat
too strong then \lam5740 does account for the 5510\ang\ absorption.
At this epoch Ti~II is blocking the blue flux even more than at day~2,
and again producing a 4060\ang\ peak.

In the day~$-3$ spectrum of SN~2000cx, the shape of the feature near
6100\ang\ is similar to its shape in SNe~1991T and 1997br (L01).  This
line shape, with its steep blue edge and nearly horizontal part
extending from about 6100\ang\ to 6300\ang, has been thought to be
odd.  But the reason the shape looks odd is that the line is rather
weak and superimposed on a strongly sloping continuum.  When the
spectrum is tilted to make the underlying continuum relatively flat,
the line profile looks like that of a normal resonance--scattering
feature.  The resemblance of this feature in SN~2000cx to the same
feature in SNe~1991T and 1997br, in which no conspicuous ``H$\beta$''
features are seen, suggests that there is little room for an H$\alpha$
contribution to the spectrum of SN~2000cx at this epoch.  But as
discussed in T03, if H$\alpha$ is confined to a clump in front of the
photosphere and its source function is higher than that of resonance
scattering as it ordinarily is in SNe~II, it may be hard to see.  (In
the special case that its source function equals the intensity from
the photosphere, the spectroscopic signature of H$\alpha$ confined to
a clump in front of the photosphere would vanish.)

\subsection{The Day $-5.5$ Infrared Spectrum}

Figure~21 shows a comparison of the short wavelength part of the
day~$-5.5$ infrared spectrum, obtained by R02, with a straightforward
extension of the optical synthetic spectrum of Figure~21, for
day~$-3$.  The PV Mg~II features in the synthetic spectrum may have
counterparts in the observed spectrum but the synthetic features are
too strong, possibly due to the mismatch of epochs if the Mg~II
optical depth is increasing with time. The most important point is
that the 1.02~$\mu$m absorption cannot be due to Mg~II \lam10926
forming above 20,000 \kms, as assumed in R02, because Mg~II \lam9226
blueshifted by 20,000 \kms\ would produce a strong absorption at
8600\ang, near a peak in the observed spectrum.  In LTE Mg~II \lam9226
has a higher optical depth than Mg~II \lam10926 for any reasonable
excitation temperature.  The Mg~II \lam 10926 and \lam9226 absorptions
do appear to be present in other SNe~Ia, but at lower velocities
ranging from about 11,000 to 15,000 \kms\ (Marion et~al. 2003).

Figure~21 also presents the same synthetic spectrum, but including
He~I lines detached at 21,000 \kms.  This shows that HV He~I \lam10830
could be responsible for at least the core of the 1.02~$\mu$m
absorption.  If the HV matter contains hydrogen then it also contains
helium, although nonthermal excitation would be required to produce a
significant optical depth in He~I \lam10830.  R02 mentioned that He~I
\lam20581 is not present in their spectra, but detailed calculations
will be required to determine whether this is inconsistent with the
presence of He~I \lam10830.

In the synthetic spectrum of Figure~21 hydrogen P$\beta$ is so weak
that it provides no evidence for or against the presence of H$\beta$
in SN~2000cx.  When extended to longer wavelengths the synthetic
spectrum for day~$-3$ does contain a weak blueshifted hydrogen
P$\alpha$ absorption component that is not seen in the observed
spectrum, but P$\alpha$, like H$\alpha$, may have an elevated source
function with respect to resonance scattering, and be difficult to see
when confined to a clump.

\section{DISCUSSION}

We have found that in the blue the early photospheric--phase spectra
of SN~2000cx are composite, containing not only relatively
high--excitation features just above the photospheric velocity (PV) of
$\sim$10,000 \kms, but also lower--excitation features forming at much
higher velocity (HV), above $\sim$20,000 \kms.  It is clear that the
HV features are partly responsible for the photometric peculiarities
of SN~2000cx.  For example, HV Ti~II features are present in the $B$
band.  For the day~$-3$, 2, 7, 15, and 32 synthetic spectra we have
used Ti~II reference--line optical depths of 0.7, 0.4, 0.6, 0.3, and
0.0, respectively, not inconsistent within our fitting freedom with a
smoothly decreasing Ti~II optical depth.  Other things being equal,
the decreasing blocking of blue flux by HV Ti~II lines would cause the
$B$ band to rise more quickly than normal, and then to decline more
slowly than normal.  This is likely to be at least partly responsible
for the lopsided $B$--band light curve.  In our present 1-D
calculations, most of the flux removed by Ti~II absorption features
tends to be re--emitted just to the red, still in the $B$ band.  But
if Ti~II lines form in a clump in front of the photosphere, then the
absorbed flux is emitted into all directions and the observer sees
strong net absorption.  Thus quantitatively evaluating the effects of
HV line formation on photometry is a task for 3-D calculations.

Similarly, the presence of HV Ca~II infrared triplet absorption in the
$I$ band may be partly responsible for the unusual behavior of the
$I$--band light curve discussed by L01 and C03.  We can see from
Figs.~1, 15, and 17 that HV Ca~II remains strong even as late as
day~32.  The evolution of the line strengths of clump--confined matter
can be complex, reflecting not only the usual Sobolev $t^{-2}$
optical--depth decline for homologous expansion, and whatever
excitation and ionization changes may occur, but also the evolving
fraction of the photosphere that is covered by the clump.  When the
photosphere recedes with respect to the homologously expanding matter
an increasing fraction of the photosphere may be covered by the clump.

The only plausible identification that we can find for the 4530\ang\
absorption is HV H$\beta$.  This identification is tentative, however,
because there is no supporting evidence, and the lack of a conspicuous
H$\alpha$ signature at day~$-3$ may require fine tuning of the
H$\alpha$ source function.

The origin of the HV matter in SN~2000cx remains unclear. If the HV
matter consists of unburned carbon and oxygen with solar abundances of
heavier elements, then the required HV mass of $\sim10^{-3}$~M$_\odot$
could be contained in a HV clump of ejecta (T03).  But if the HV
matter is hydrogen--rich then it must have come from a nondegenerate
companion star (Marietta, Burrows, \& Fryxell 2000), because
$10^{-3}$~M$_\odot$ of hydrogen could not have been present on the
surface of the progenitor white dwarf without burning.  From the
absence of a narrow circumstellar H$\alpha$ feature in early
high--resolution spectra, Lundqvist et~al. (2003) infer an upper limit
to the wind mass--loss rate of the companion of $\sim10^{-5}$
M$_\odot$~yr$^{-1}$.

The PV spectra of SN~2000cx also are peculiar.  After allowing for the
now--recognized role of HV Ti~II features in the blue, the case for a
spectroscopic relation between the underlying PV spectra of SNe~2000cx
and 1991T is perhaps strengthened.  We confirm and extend the findings
of L01 that Fe~III lines persisted for an unusually long time after
maximum, that line blueshifts were higher than normal, and that Fe~II
lines were late to develop.  We find that when Fe~II lines finally did
develop during the late photospheric phase, they formed out to higher
velocities than in normal SNe~Ia.  All of this seems consistent with
the suggestion by L01 that SN~2000cx was a more powerful version of
SN~1991T, but as emphasized by C03, a reliable distance to NGC~524 is
needed to determine the luminosity of SN~2000cx.

Although HV line formation is unusually strong in SN~2000cx, it has
become clear that it sometimes occurs in other SNe~Ia.  HV Ca~II and
Fe~II were first recognized in premaximum spectra of SN~1994D (Hatano
et~al. 1999a); strong signatures of HV Ca~II have been detected in
premaximum flux and polarization spectra of SN~2001el (Wang
et~al. 2003; Kasen et~al. 2003); HV Ca~II can be seen in premaximum
spectra of SNe~1999ee (Hamuy et~al. 2002; P.~Mazzali, in preparation);
and HV Fe~II has been invoked to fit both premaximum and postmaximum
spectra of SN~1998aq (B03).  Recently Gerardy et~al. (2003) have
attributed HV Ca~II in SN~2003du to a dense shell formed by
interaction of the high--velocity ejecta with $\sim2 \times 10^{-2}$
M$_\odot$ of hydrogen--rich circumstellar matter.  Further close
scrutiny of SN~Ia spectra to look for more subtle signs of HV line
formation is needed.  An indication of which ions would be likely to
appear at various temperatures, for carbon--oxygen rich and hydrogen
rich compositions, can be obtained from Figs. 11 and 12 of TO3.  These
figures show that (in LTE) when H, Ca~II, Ti~II, and Fe~II are
present, additional possibilities include Na~I, Fe~I, Sr~II, and
Ba~II, depending on the temperature and electron density.

The ``H$\beta$'' feature of SN~2000cx has never been seen, or at least
never recognized, in other SNe~Ia.  If the HV H$\beta$ identification
in SN~2000cx is correct, and yet HV H$\beta$ seldom if ever appears in
the spectra of other SNe~Ia, even those that exhibit HV Ca~II
features, one reason could be that the outer layers of most donor
stars in SN~Ia binary progenitor systems are helium--rich rather than
hydrogen--rich (e.g., Kato \& Hachisu 2003 and references therein).

Hamuy et~al. (2003) have discovered a hydrogen signature in SN~2002ic
that differs from the putative clump--confined HV hydrogen associated
with SN~2000cx.  The H$\alpha$ emission associated with SN~2002ic has
a characteristic velocity of only $\sim1800$ \kms.  It is unlikely
that this kind of circumstellar hydrogen signature would have gone
unrecognized in previously observed SNe~Ia, so it appears that
SN~2002ic, although very interesting, is a special case. 

This work has been supported by NSF grants AST-9986965, AST-0204771,
and AST-0307323, NASA grant NAG 5-12127, and grant HST-AR-09544-01A
(provided by NASA through the STScI, operated by AURA Incorporated,
under NASA contract NAS5-26555). RJR was supported by The Aerospace
Corporation's Independent Research and Development Program.

\clearpage     

\begin{figure}
\includegraphics[width=.7\textwidth,angle=270]{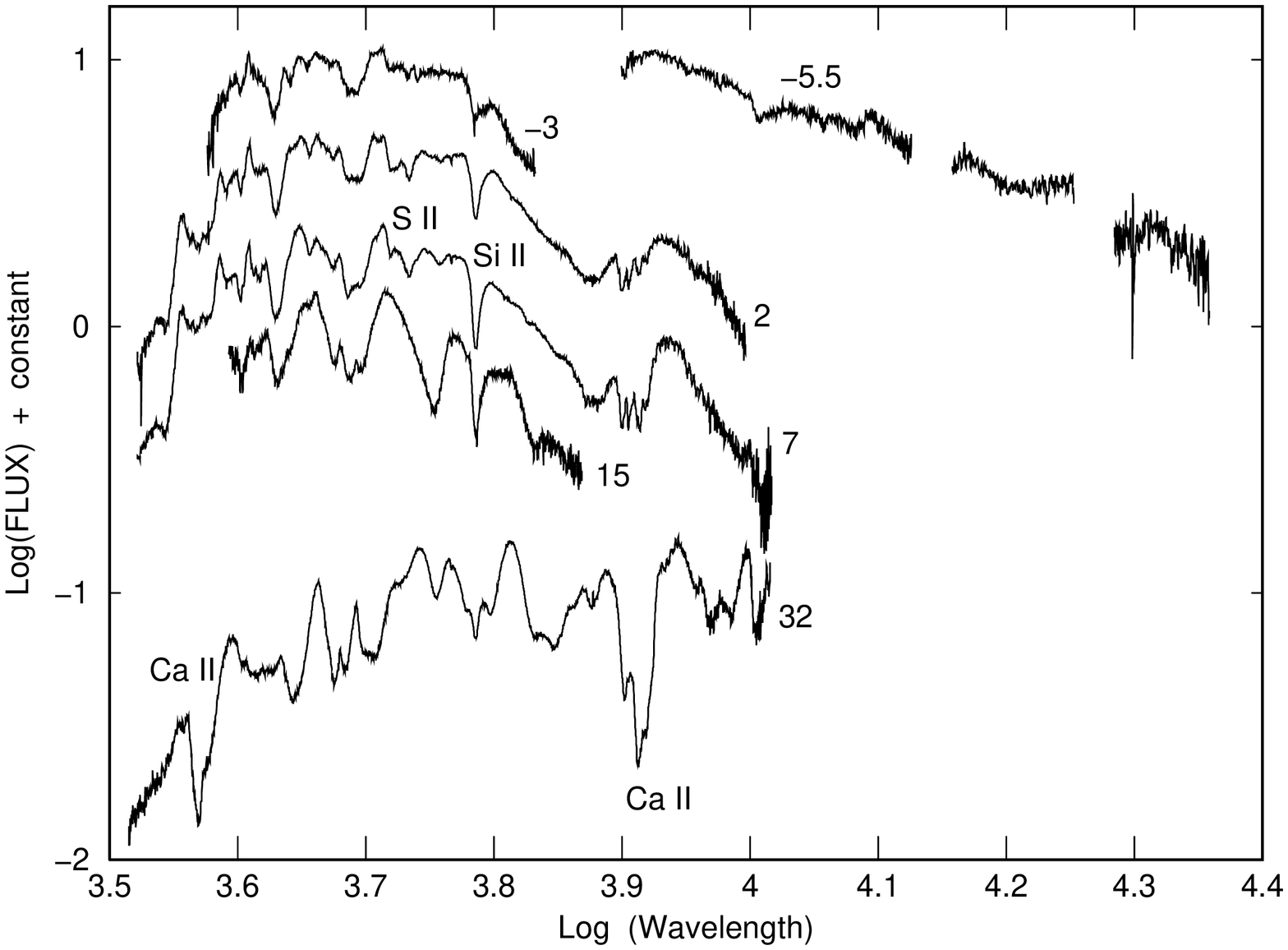}
\caption{Five optical spectra of SN~2000cx from L01 and an infrared
  spectrum from R02 are displayed.  Epochs are in days with respect to
  the date of maximum brightness in the $B$ band, 2000 July~27 (L01).
  The IR spectrum is the sum of spectra obtained 5 and 6 days before
  maximum; in regions of very strong telluric absorption the IR
  spectrum is not plotted, and the noise near log($\lambda)$=4.3 is an
  artifact of the correction for telluric CO$_2$ absorption.  The flux
  of all spectra is per unit frequency interval and the vertical
  displacement is arbitrary.  All spectra shown in this article have
  been corrected for the redshift of NGC~524, $z=0.008$.  No
  correction for interstellar reddening has been applied.}
\end{figure}

\begin{figure}
\includegraphics[width=.8\textwidth,angle=270]{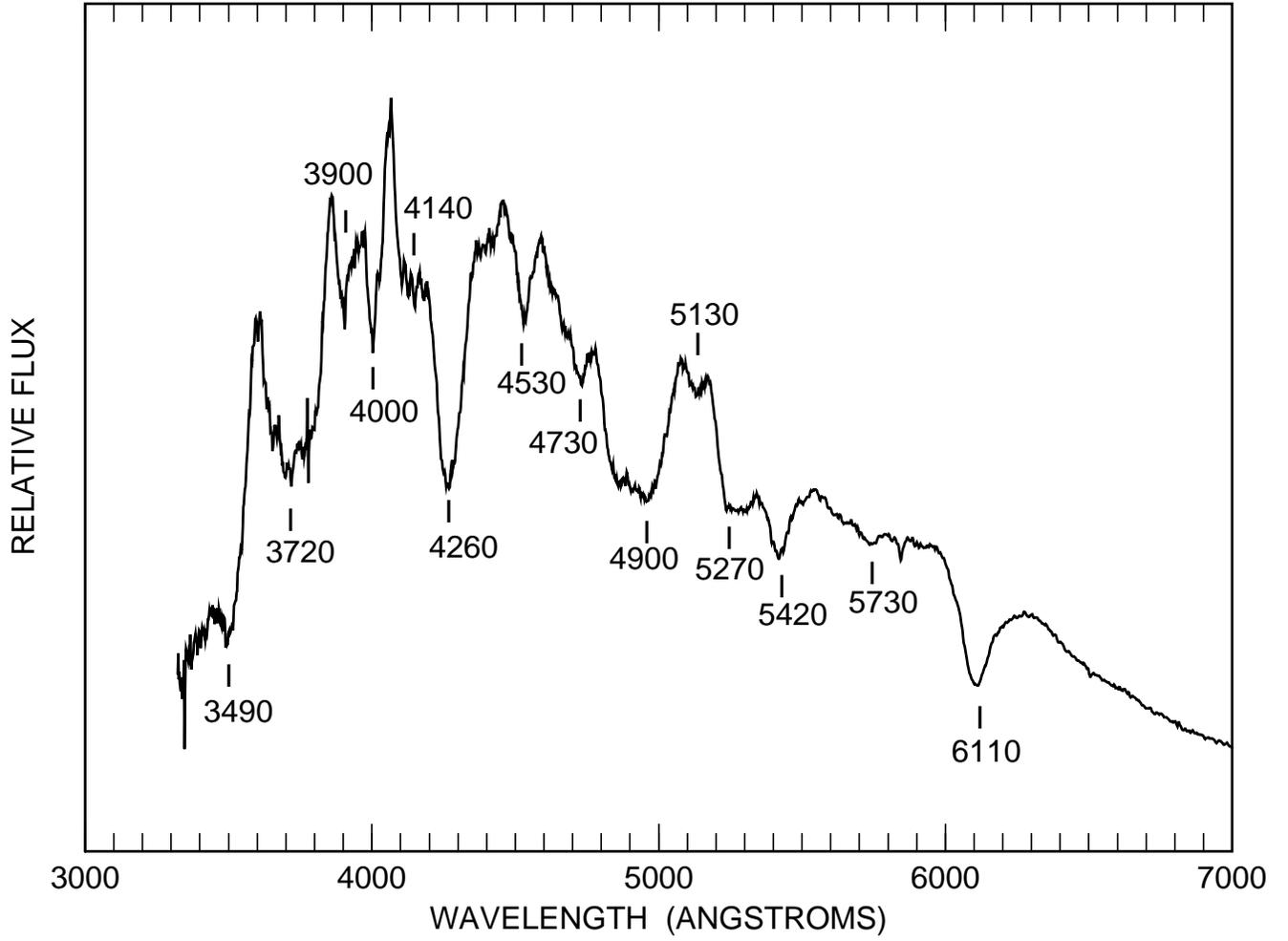}
\caption{Wavelengths of absorption features in the day~2 spectrum of
  SN~2000cx are labelled. In this and subsequent figures the flux is
  per unit wavelength interval.}
\end{figure}

\begin{figure}
\includegraphics[width=.8\textwidth,angle=270]{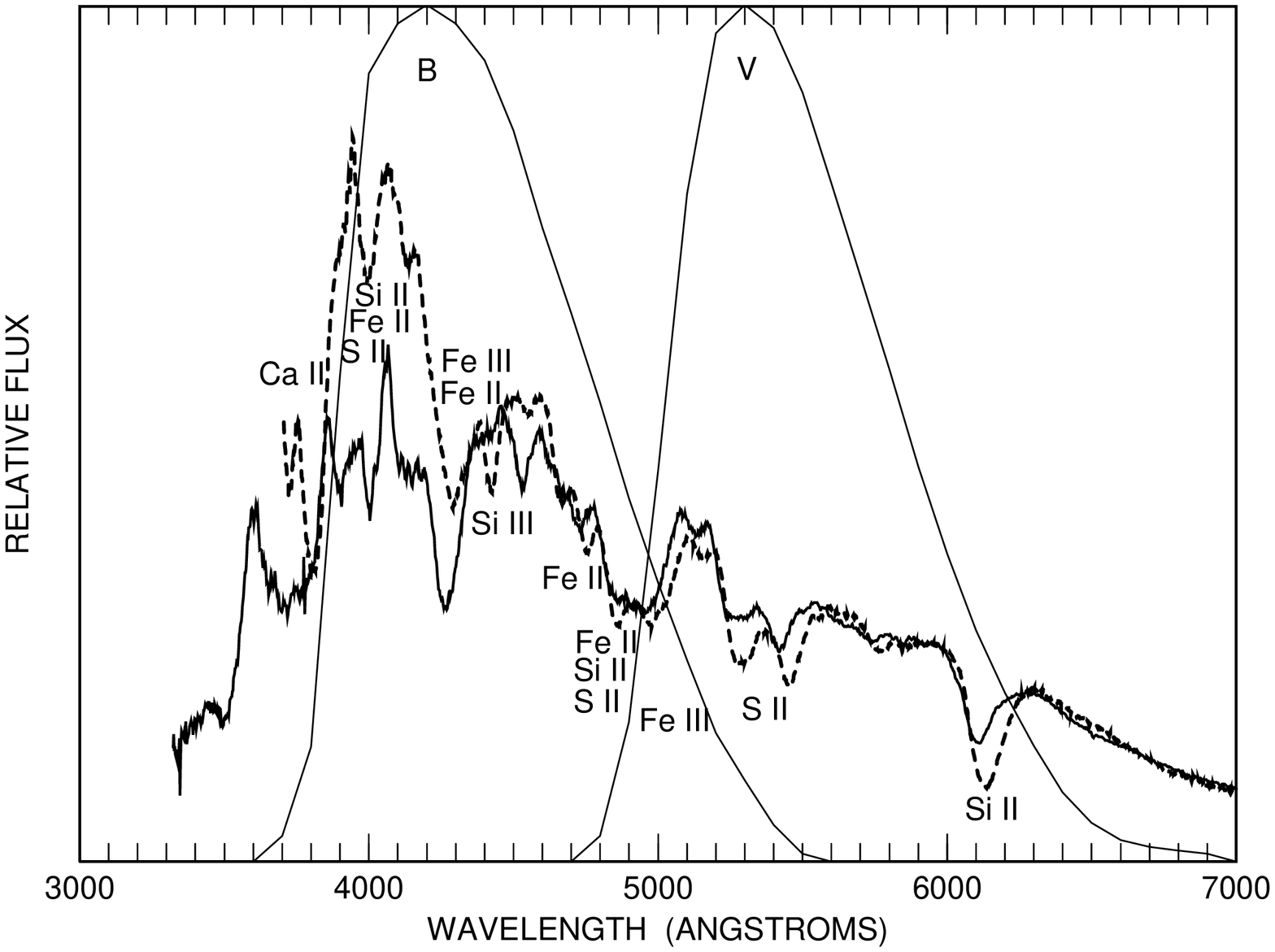}
\caption{The day~2 spectrum of SN~2000cx ({\sl thick solid line}) is
  compared with a day~2 spectrum of the normal SN~Ia 1998aq ({\sl
  dashed line}).  Ions used by B03 to account for the absorption
  features of SN~1998aq are indicated.  The $B$--band and $V$--band
  filter functions are also shown ({\sl thin solid lines}).}
\end{figure}

\begin{figure}
\includegraphics[width=.8\textwidth,angle=270]{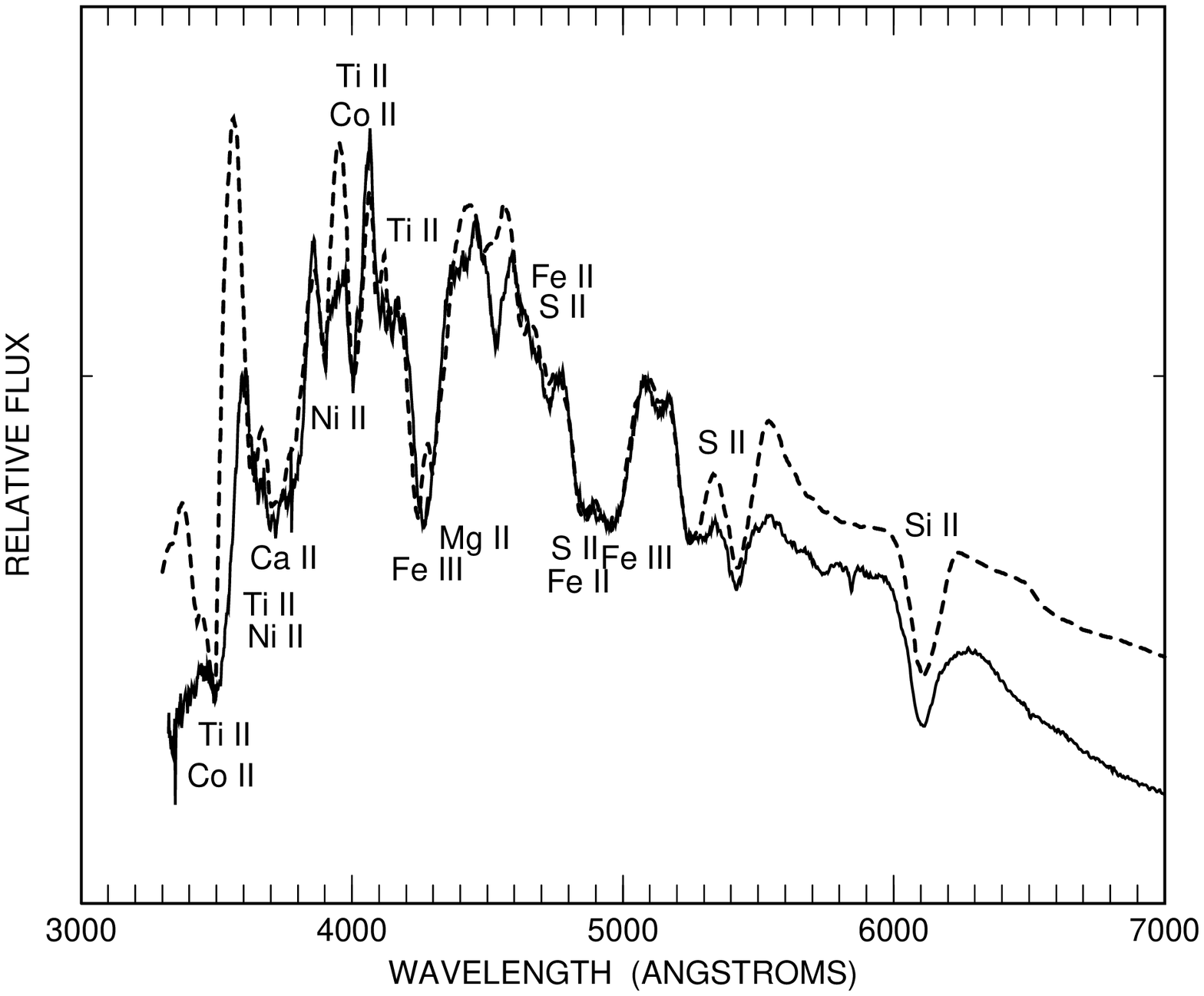}
\caption{The day~2 spectrum of SN~2000cx ({\sl solid line}) is
  compared with a synthetic spectrum ({\sl dashed line}) that has
  $v_{phot}=11,000$ \kms, $T_{bb}=9000$~K, and contains lines of nine
  ions.}
\end{figure}

\begin{figure}
\includegraphics[width=.8\textwidth,angle=270]{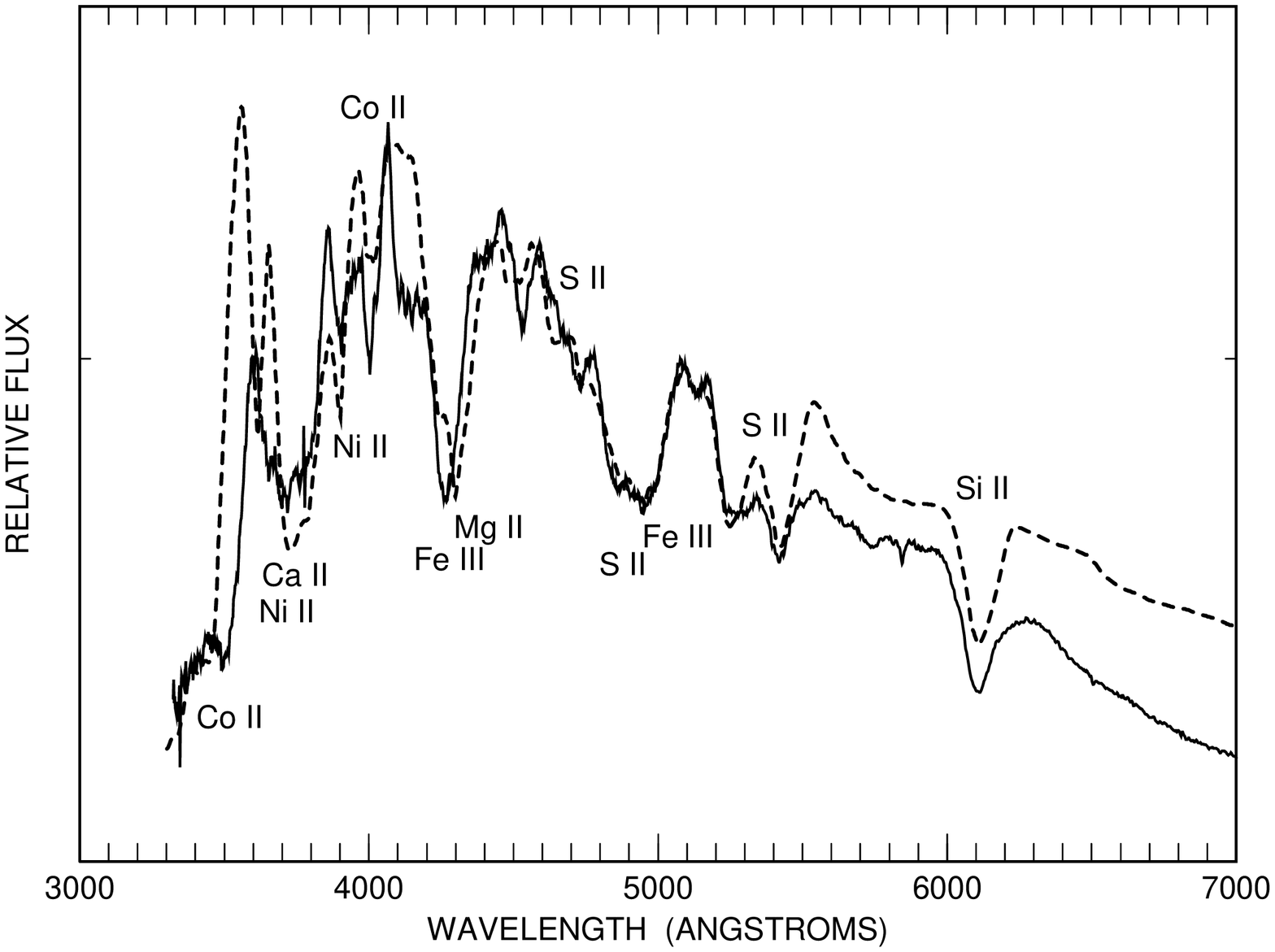}
\caption{Like Figure~4 but the synthetic spectrum includes only PV line
  formation.}
\end{figure}

\clearpage

\begin{figure}
\includegraphics[width=.8\textwidth,angle=270]{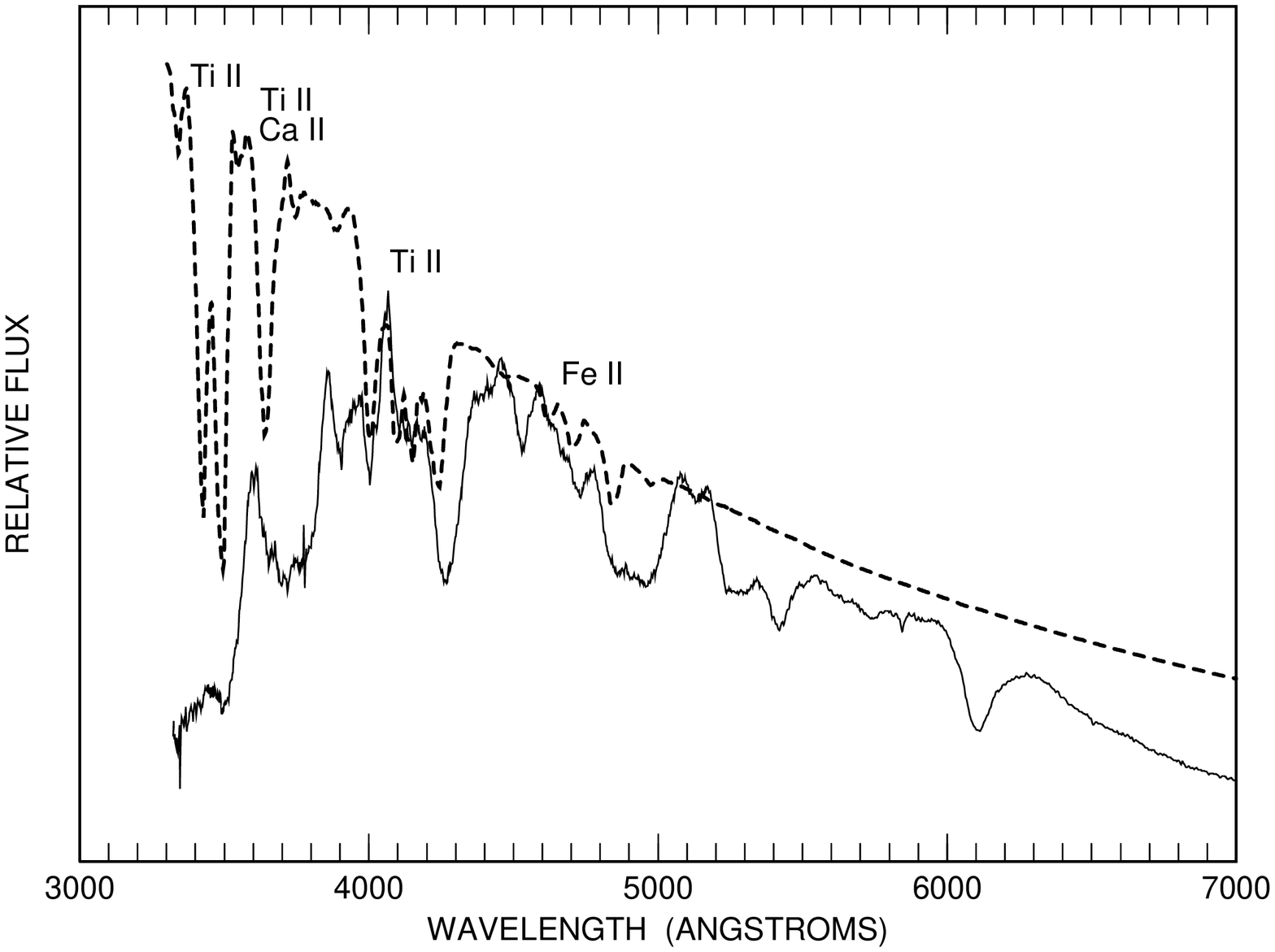}
\caption{Like Figure~4 but the synthetic spectrum includes only HV line
  formation.}
\end{figure}

\begin{figure}
\includegraphics[width=.8\textwidth,angle=270]{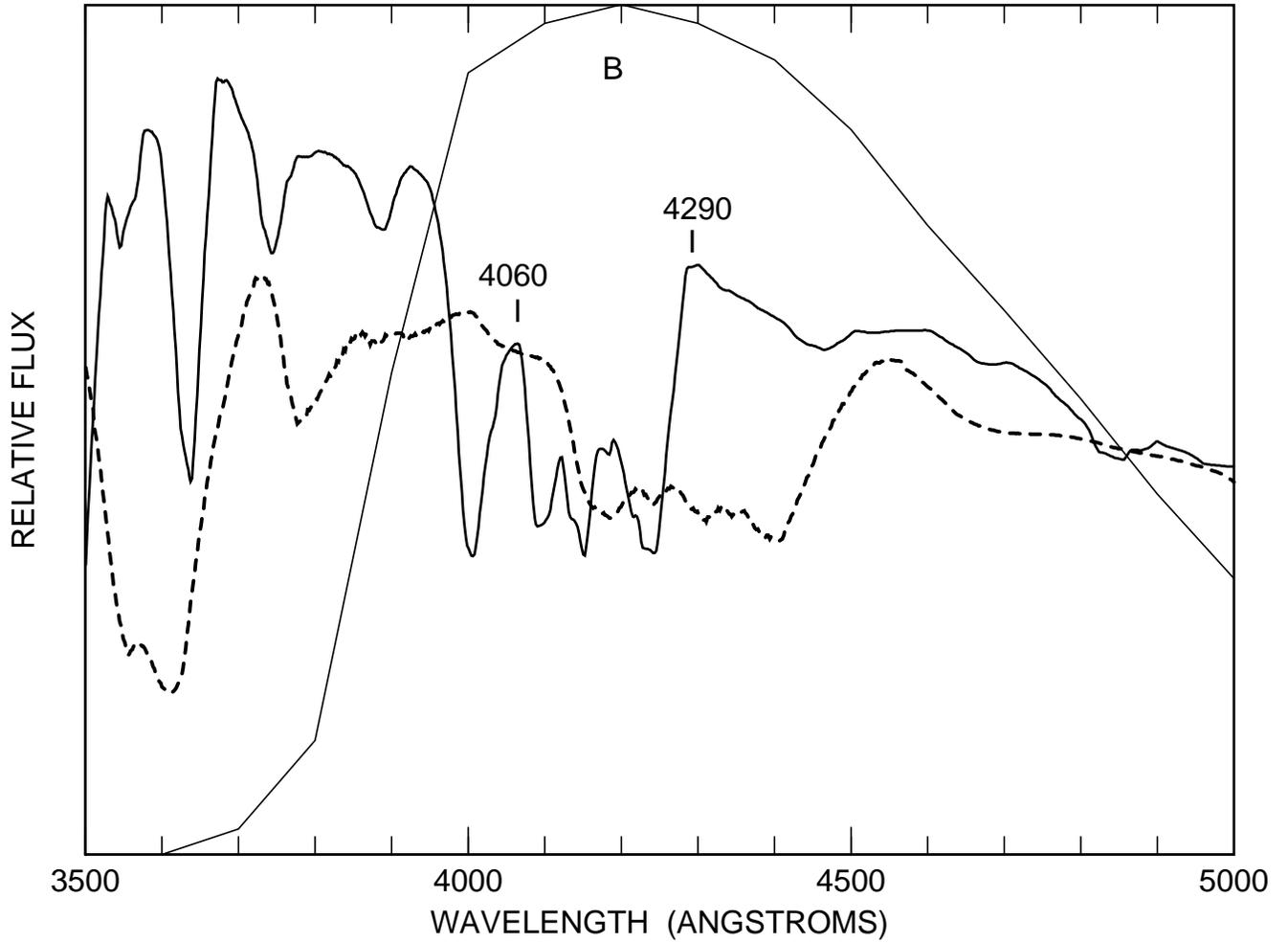}
\caption{A synthetic spectrum having $v_{phot}=11,000$ \kms\ and
  containing only Ti~II lines detached at 23,000 \kms\ ({\sl solid
  line}) is compared with a synthetic spectrum having
  $v_{phot}=11,000$ \kms\ and containing only undetached Ti~II lines
  ({\sl dashed line}). The $B$--band filter function is also shown
  ({\sl thin solid line}).}
\end{figure}

\begin{figure}
\includegraphics[width=.8\textwidth,angle=270]{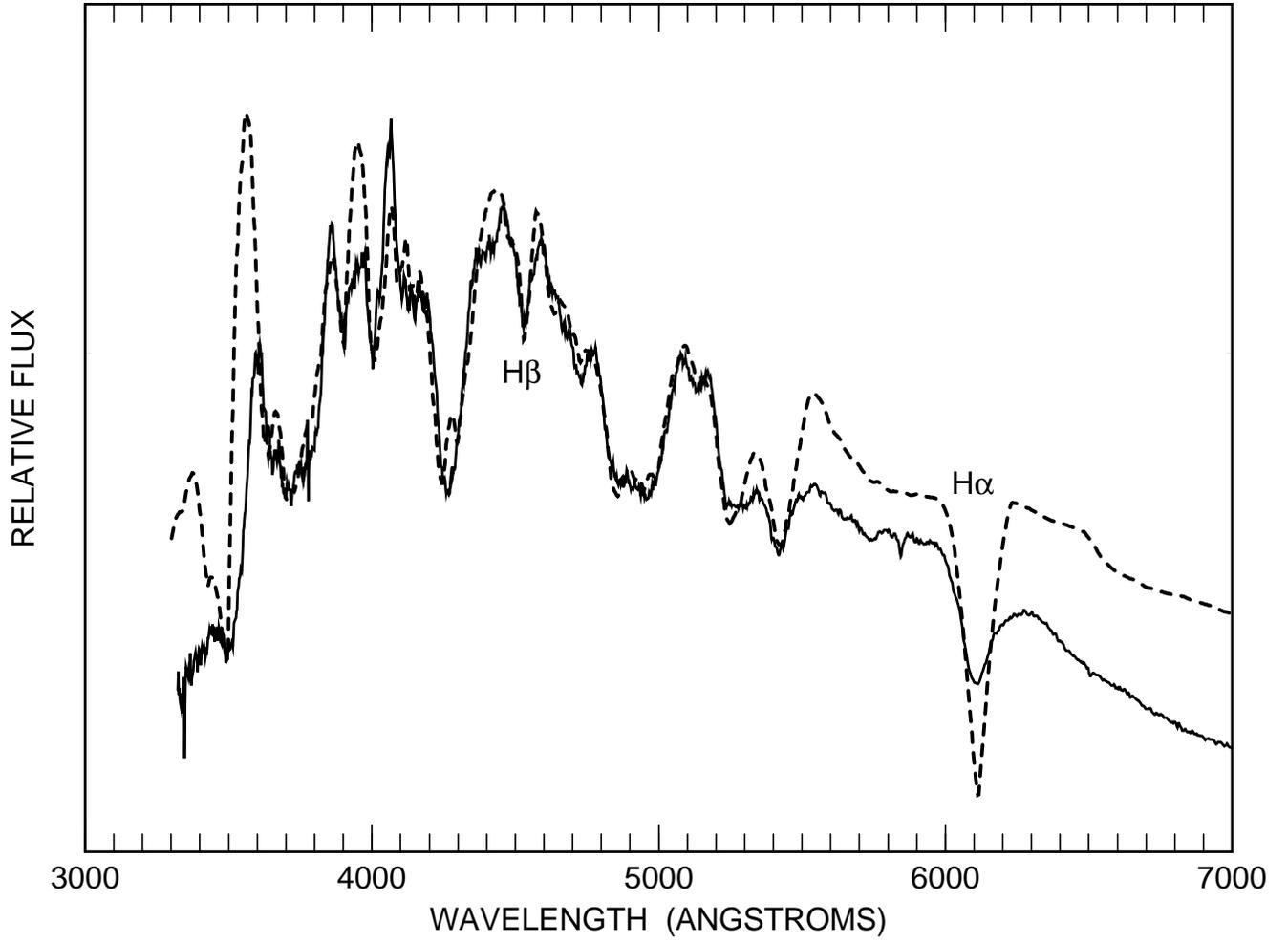}
\caption{Like Figure~4 but with hydrogen lines, detached at 22,000 \kms,
  in the synthetic spectrum.}
\end{figure}

\begin{figure}
\includegraphics[width=.8\textwidth,angle=270]{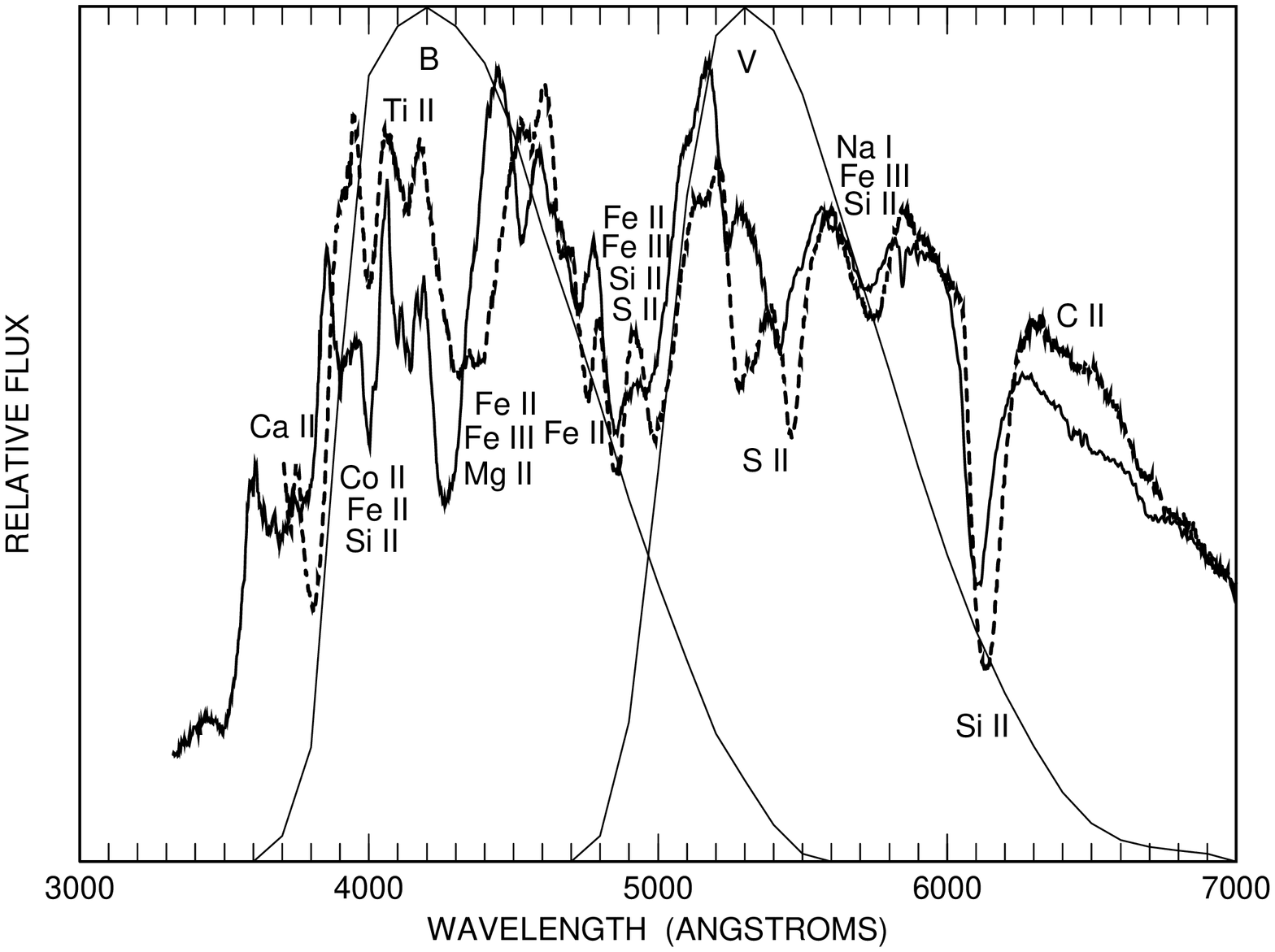}
\caption{The day~7 spectrum of SN~2000cx ({\sl thick solid line}) is
  compared with a day~7 spectrum of the normal SN~Ia 1998aq ({\sl
  dashed line}).  Ions used by B03 to account for the absorption
  features of SN~1998aq are indicated. The $B$--band and $V$--band
  filter functions are also shown ({\sl thin solid lines}).}
\end{figure}

\begin{figure}
\includegraphics[width=.8\textwidth,angle=270]{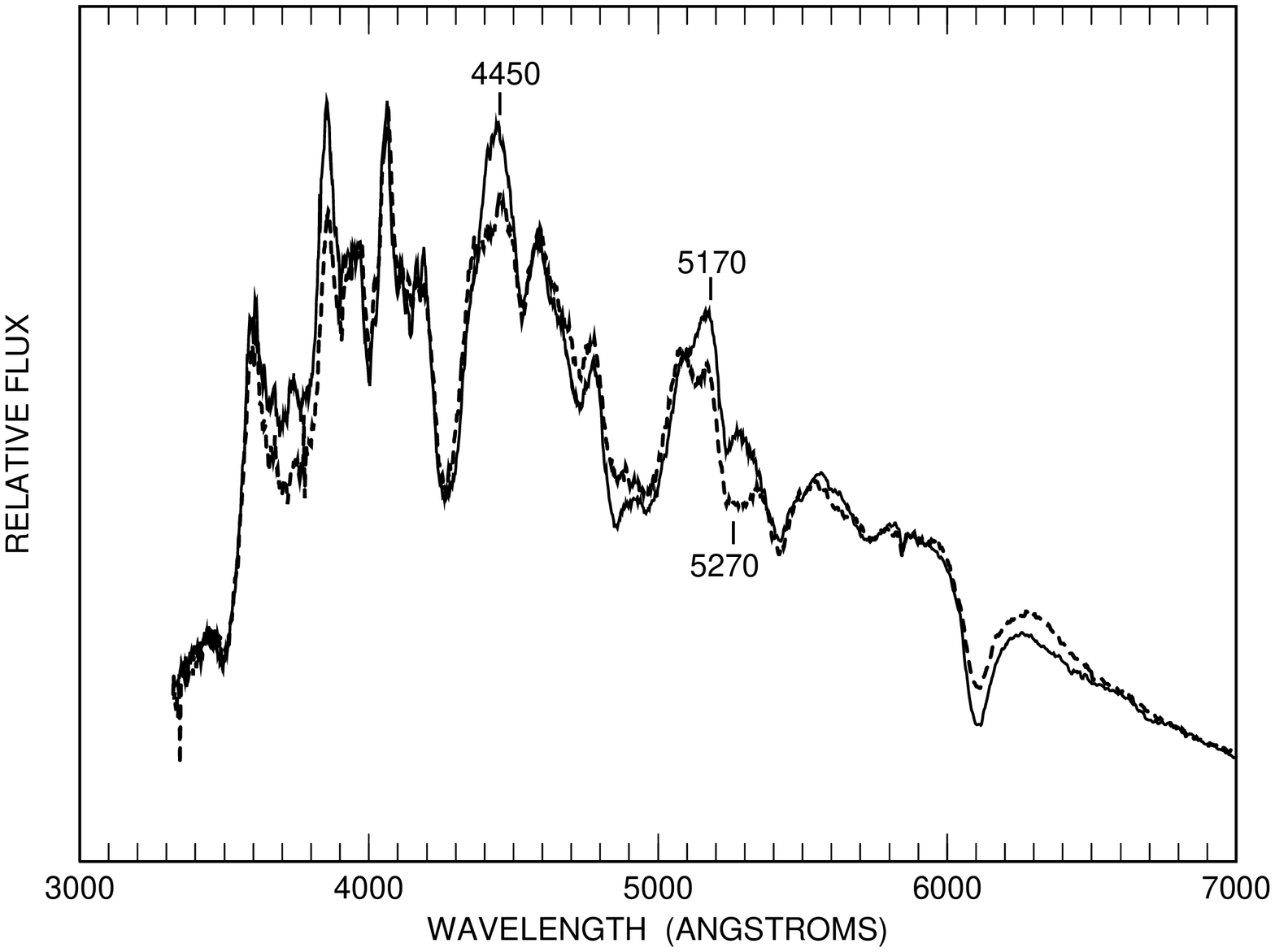}
\caption{The day~7 ({\sl solid line}) and day~2 ({\sl dashed line}) spectra of
  SN~2000cx are compared.}
\end{figure}

\begin{figure}
\includegraphics[width=.8\textwidth,angle=270]{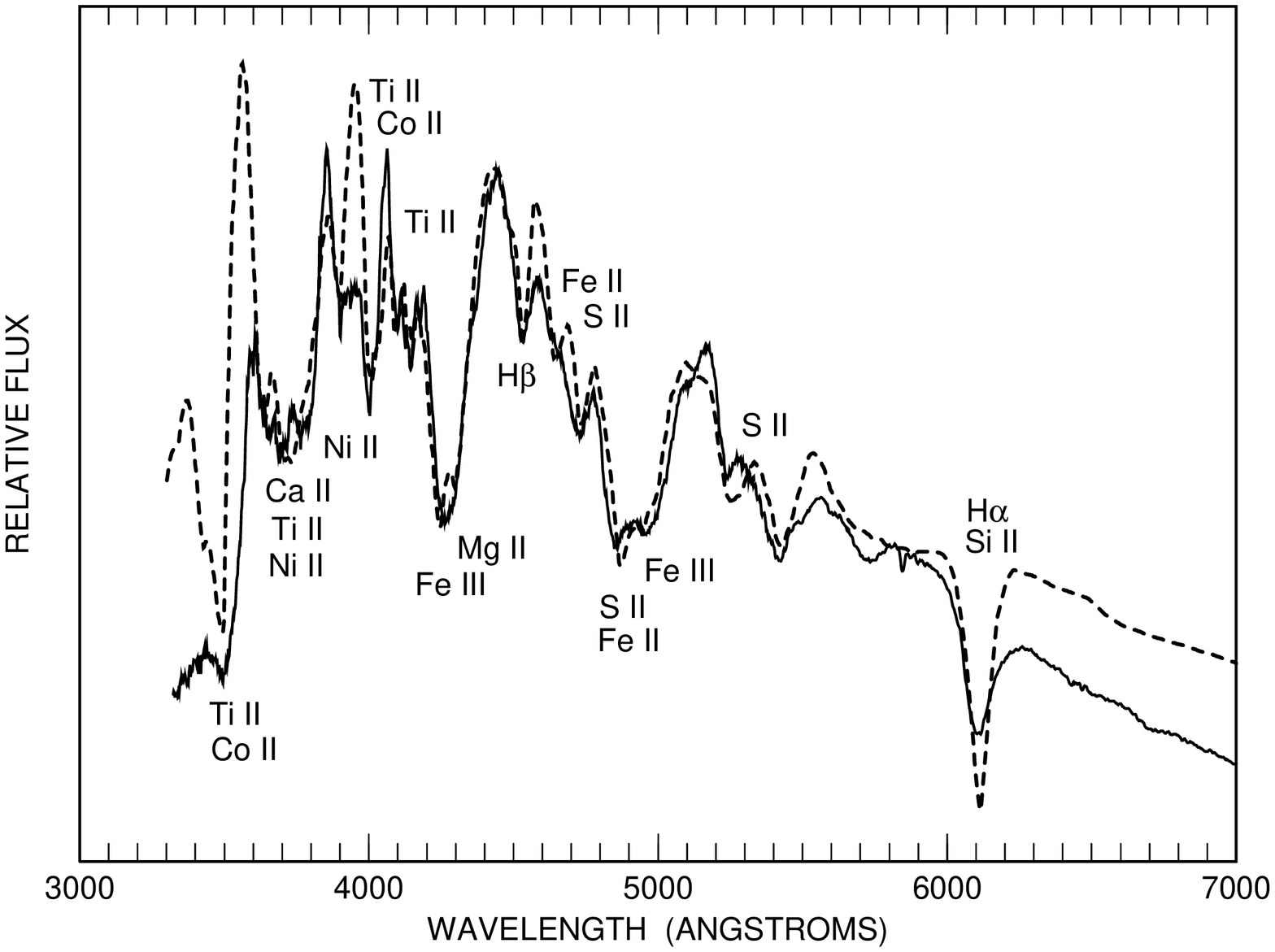}
\caption{The day~7 spectrum of SN~2000cx ({\sl solid line}) is
  compared with a synthetic spectrum ({\sl dashed line}) that has
  $v_{phot}=11,000$ \kms, $T_{bb}=11000$~K, and contains lines of nine
  ions.}
\end{figure}

\begin{figure}
\includegraphics[width=.8\textwidth,angle=270]{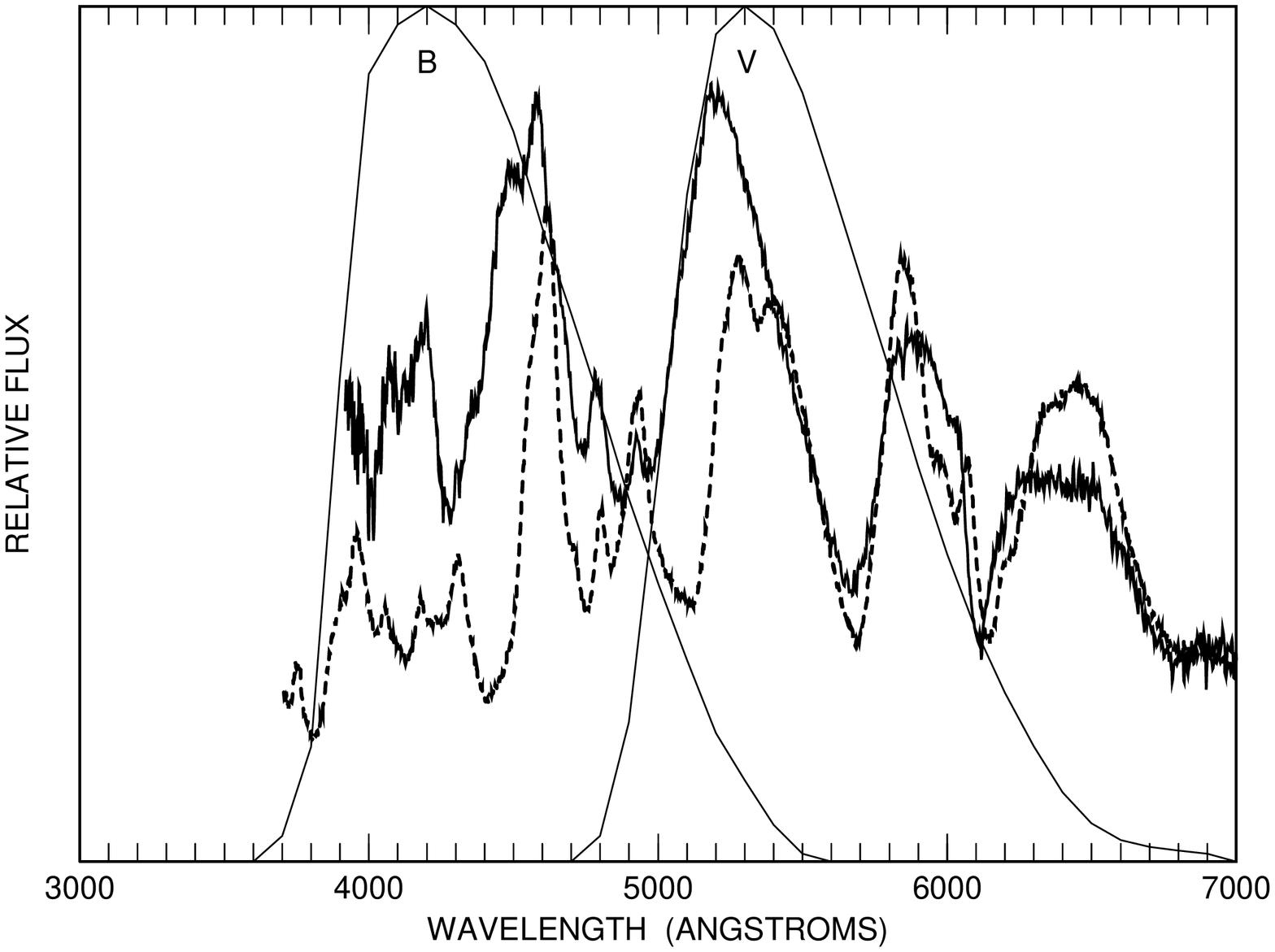}
\caption{The day~15 spectrum of SN~2000cx ({\sl thick solid line}) is
  compared with a day~18 spectrum of the normal SN~Ia 1998aq ({\sl
  dashed line}).  The $B$--band and $V$--band filter functions are
  also shown ({\sl thin solid lines}).}
\end{figure}

\begin{figure}
\includegraphics[width=.8\textwidth,angle=270]{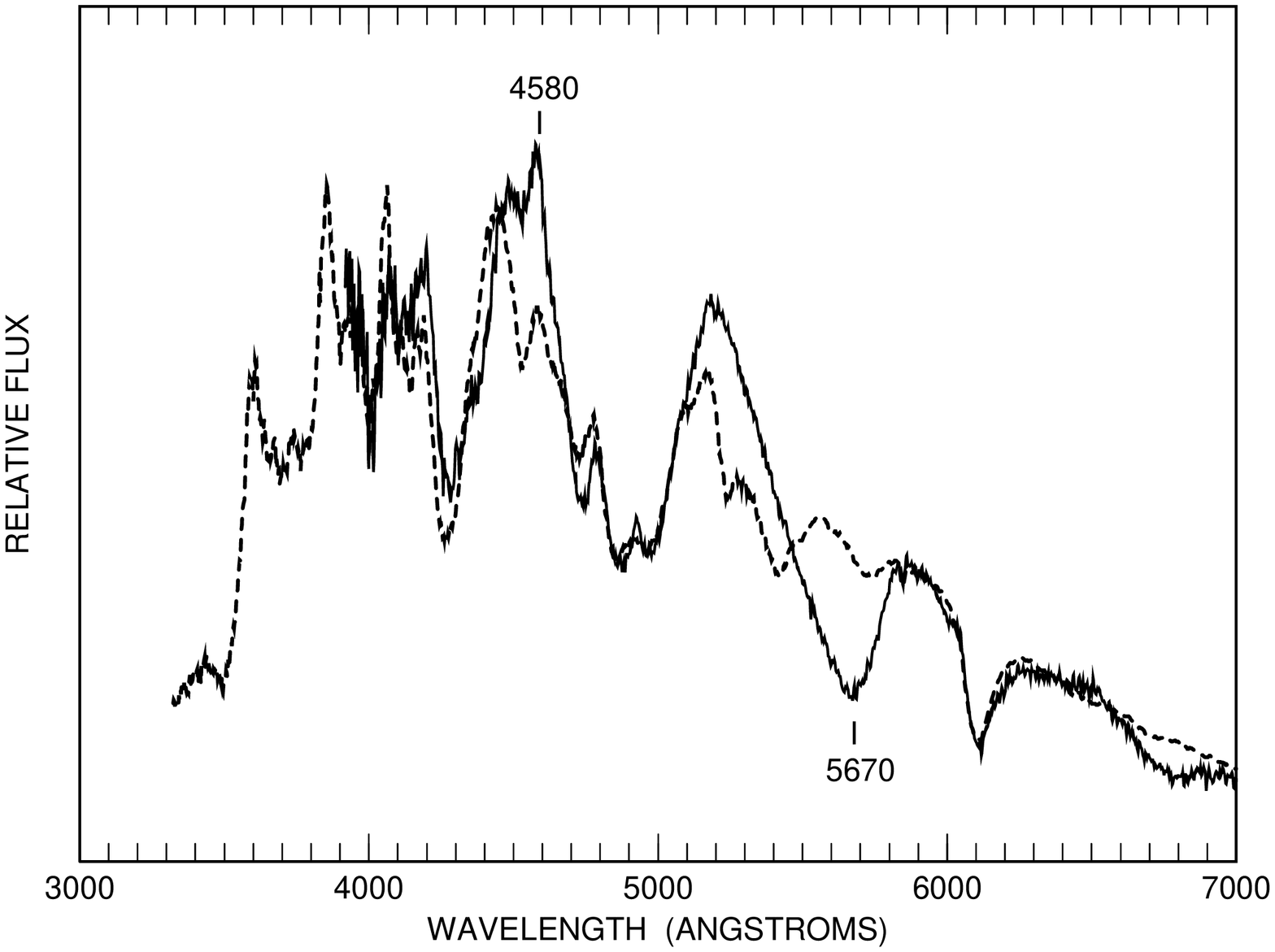}
\caption{The day~15 ({\sl solid line}) and day~7 ({\sl dashed line})
  spectra of SN~2000cx are compared.}
\end{figure}

\begin{figure}
\includegraphics[width=.8\textwidth,angle=270]{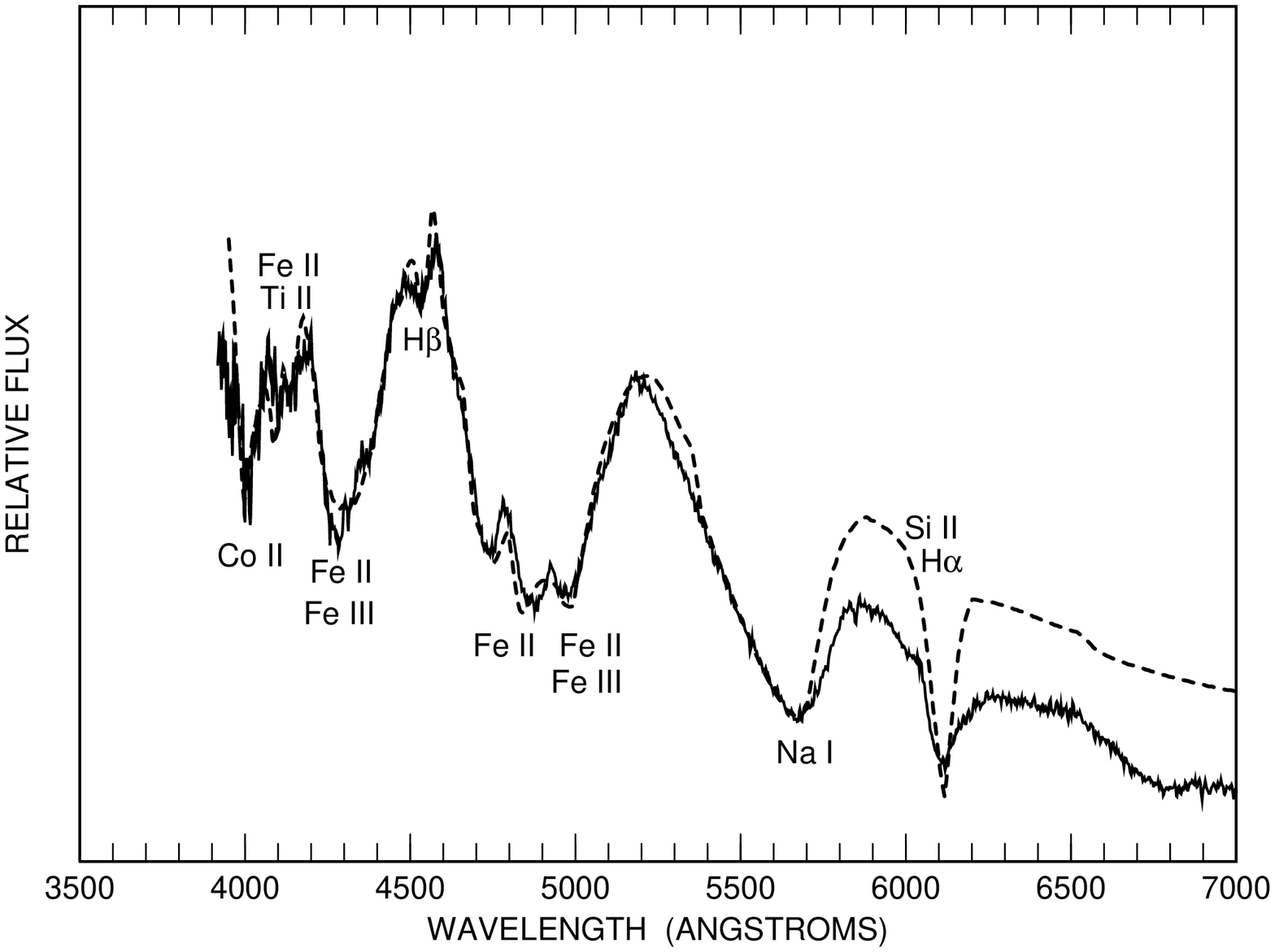}
\caption{The day~15 spectrum of SN~2000cx ({\sl solid line}) is
  compared with a synthetic spectrum ({\sl dashed line}) that has
  $v_{phot}=10,000$ \kms, $T_{bb}=12000$~K, and contains lines of five
  ions.}
\end{figure}

\begin{figure}
\includegraphics[width=.8\textwidth,angle=270]{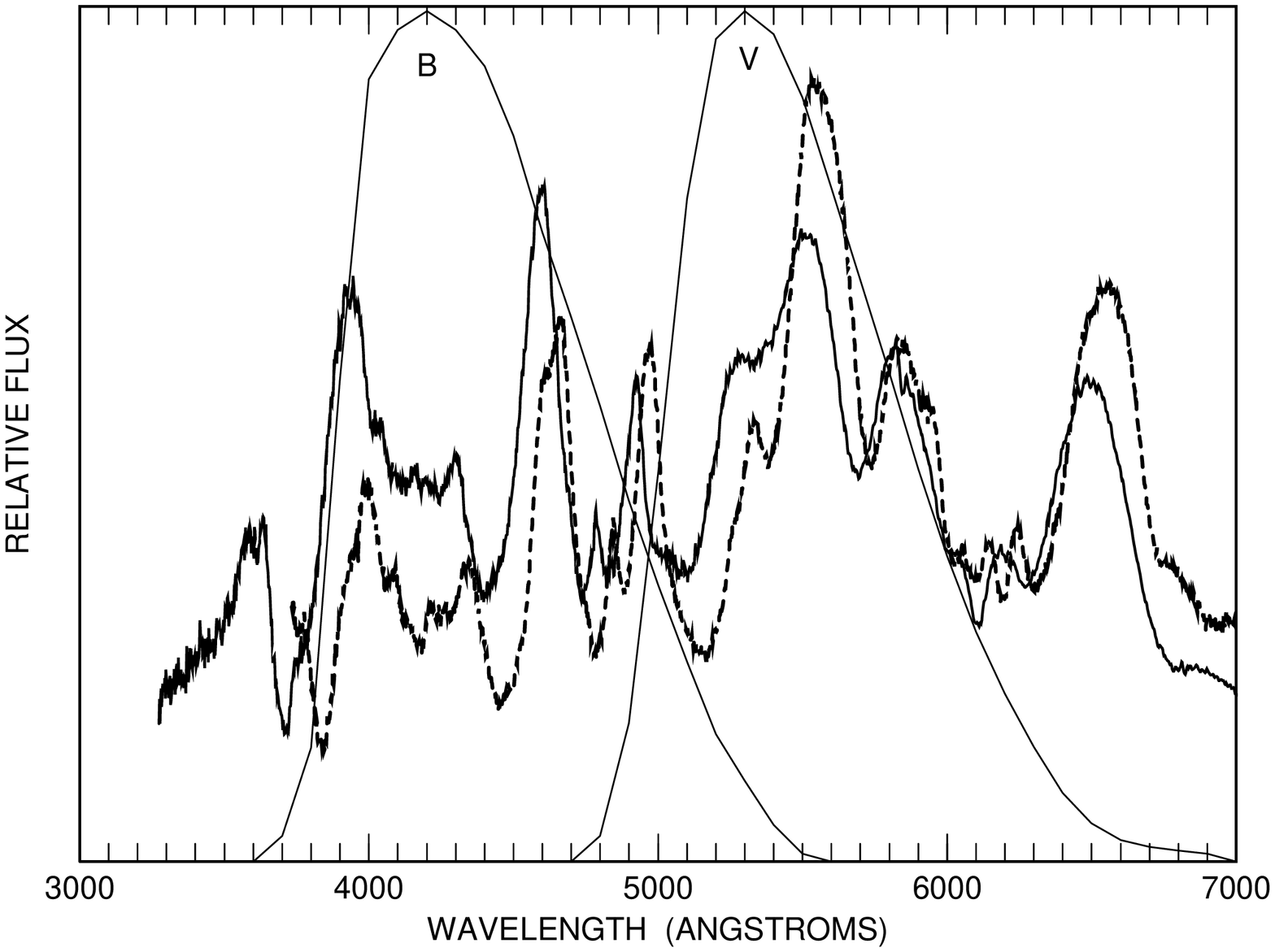}
\caption{The day~32 spectrum of SN~2000cx ({\sl thick solid line}) is
  compared with a day~32 spectrum of the normal SN~Ia 1998aq ({\sl
  dashed line}).  The $B$--band and $V$--band filter functions are
  also shown ({\sl thin solid lines}).}
\end{figure}

\begin{figure}
\includegraphics[width=.8\textwidth,angle=270]{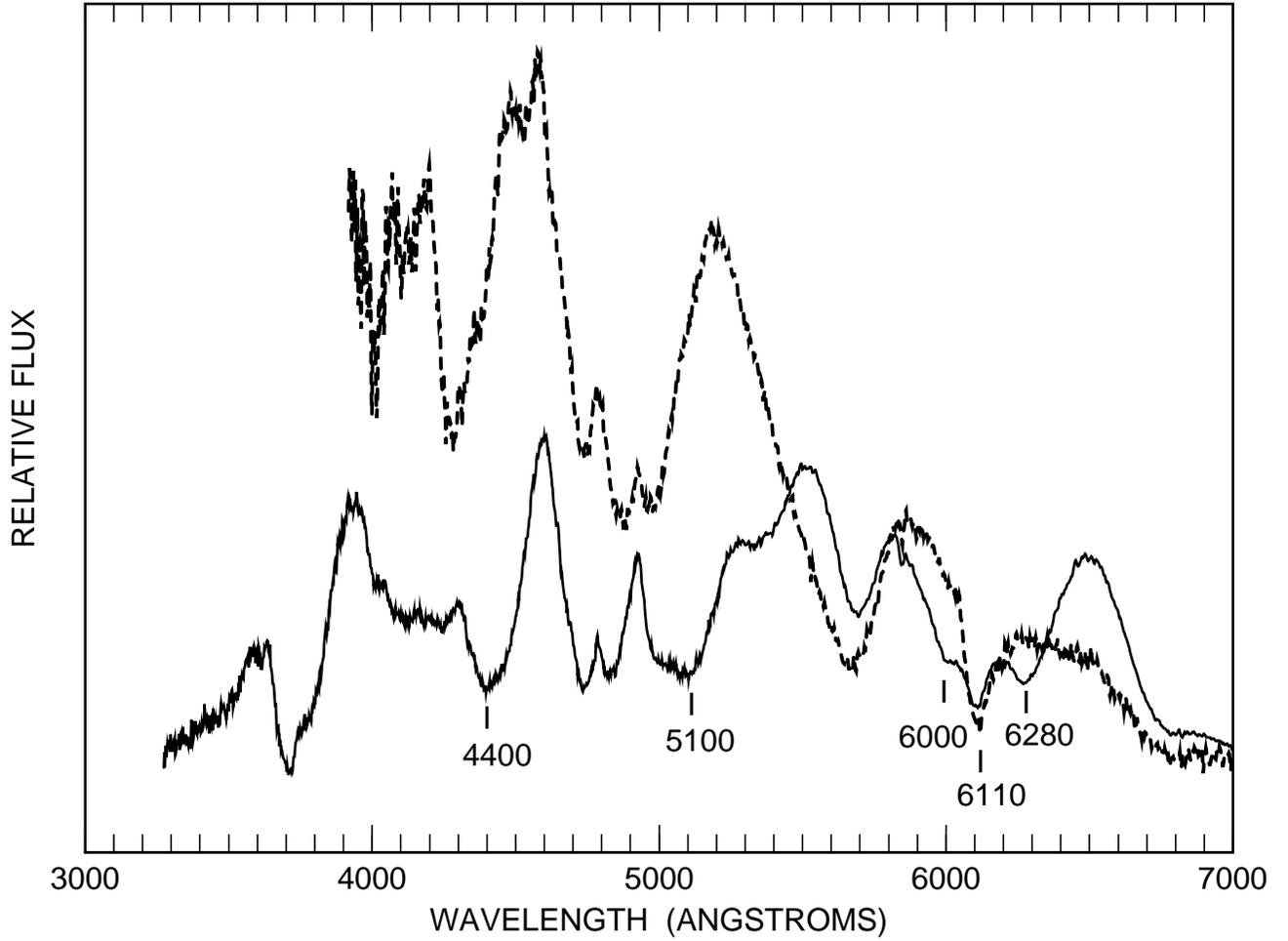}
\caption{The day~32 ({\sl solid line}) and day~15 spectra ({\sl dashed
  line}) of SN~2000cx are compared.}
\end{figure}

\begin{figure}
\includegraphics[width=.8\textwidth,angle=270]{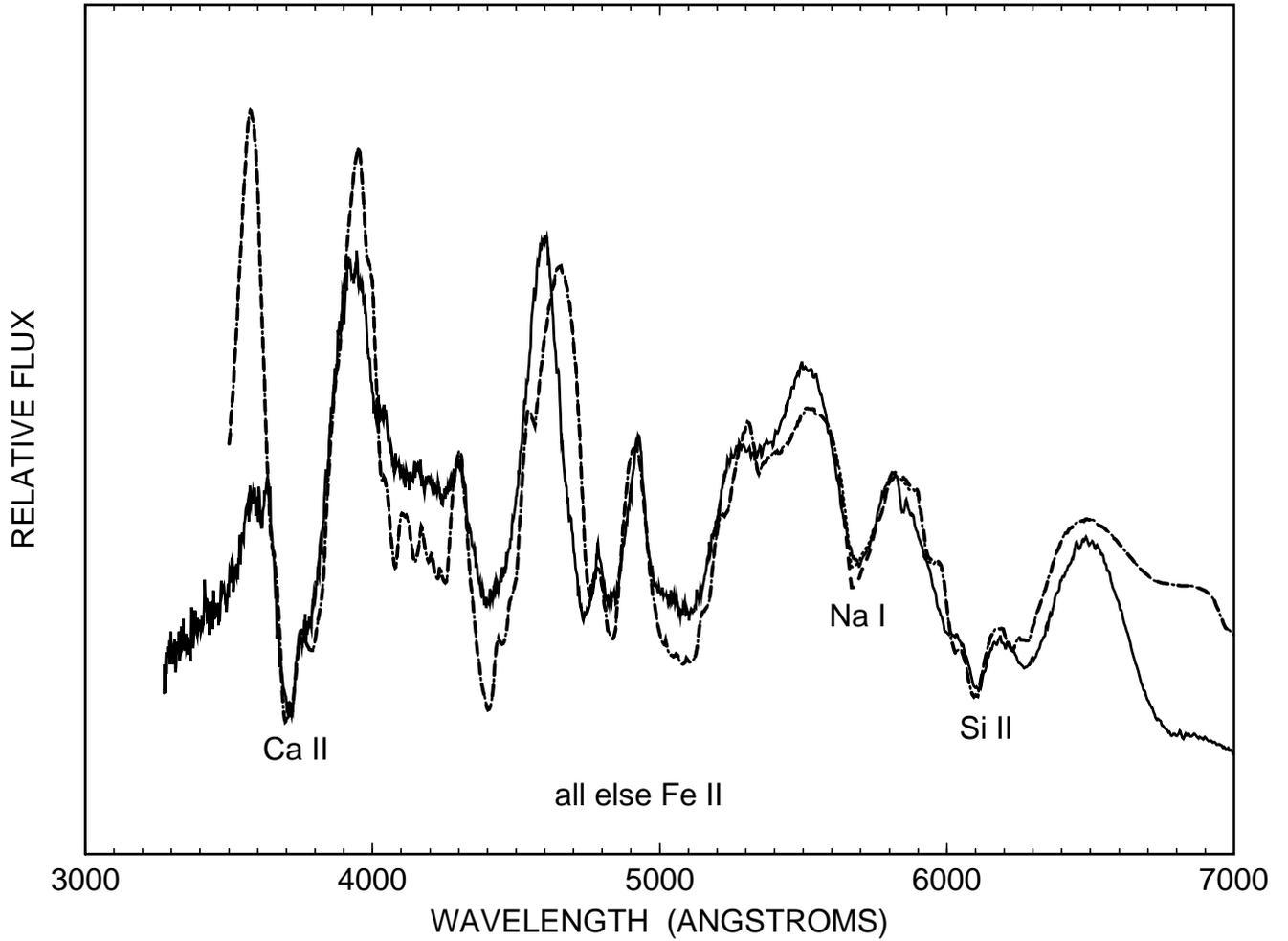}
\caption{The day~32 spectrum of SN~2000cx ({\sl solid line}) is
  compared with a synthetic spectrum ({\sl dashed line}) that has
  $v_{phot}=11,000$ \kms, $T_{bb}=8000$~K, and contains lines of four
  ions.  All unlabelled features in the synthetic spectrum are
  produced by Fe~II.}
\end{figure}

\begin{figure}
\includegraphics[width=.8\textwidth,angle=270]{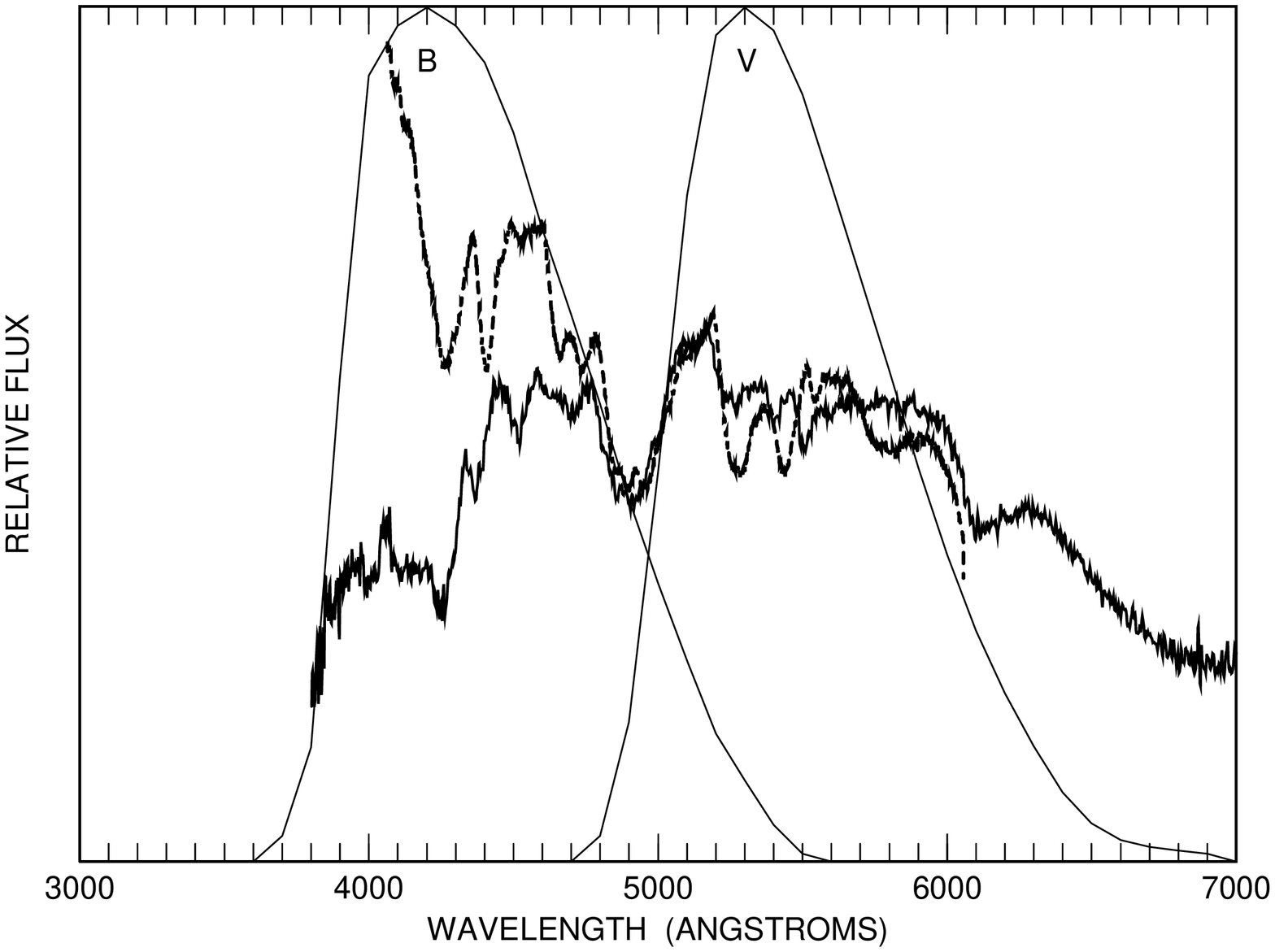}
\caption{The day~$-3$ spectrum of SN~2000cx ({\sl solid line}) is
  compared with a day~$-3$ spectrum of the normal SN~Ia 1998aq ({\sl
  dashed line}).  The $B$--band and $V$--band filter functions are
  also shown ({\sl thin solid lines}).}
\end{figure}

\begin{figure}
\includegraphics[width=.8\textwidth,angle=270]{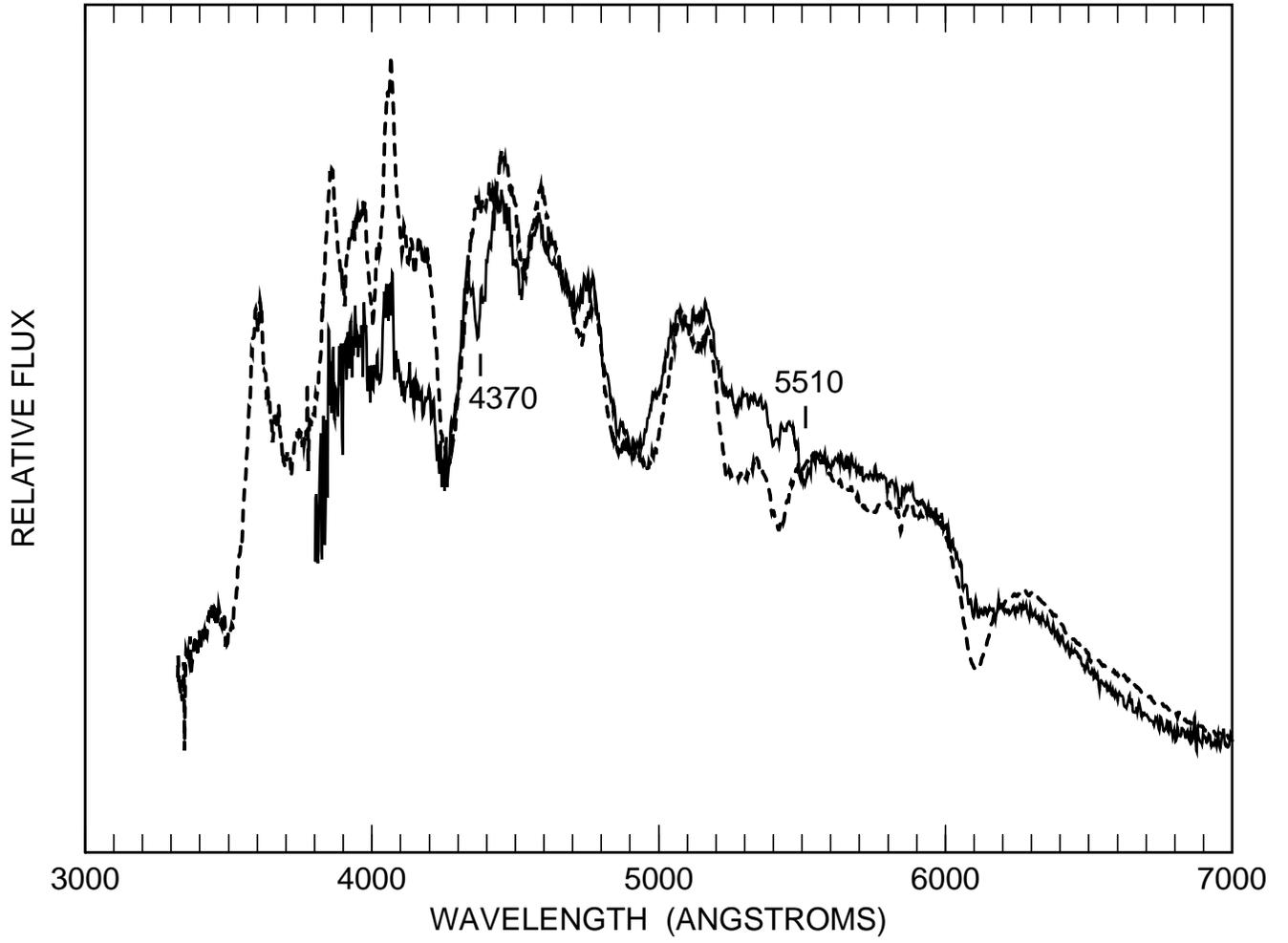}
\caption{The day~$-03$ ({\sl thick solid line}) and day~2 ({\sl dashed
  line}) spectra of SN~2000cx are compared.}
\end{figure}

\begin{figure}
\includegraphics[width=.8\textwidth,angle=270]{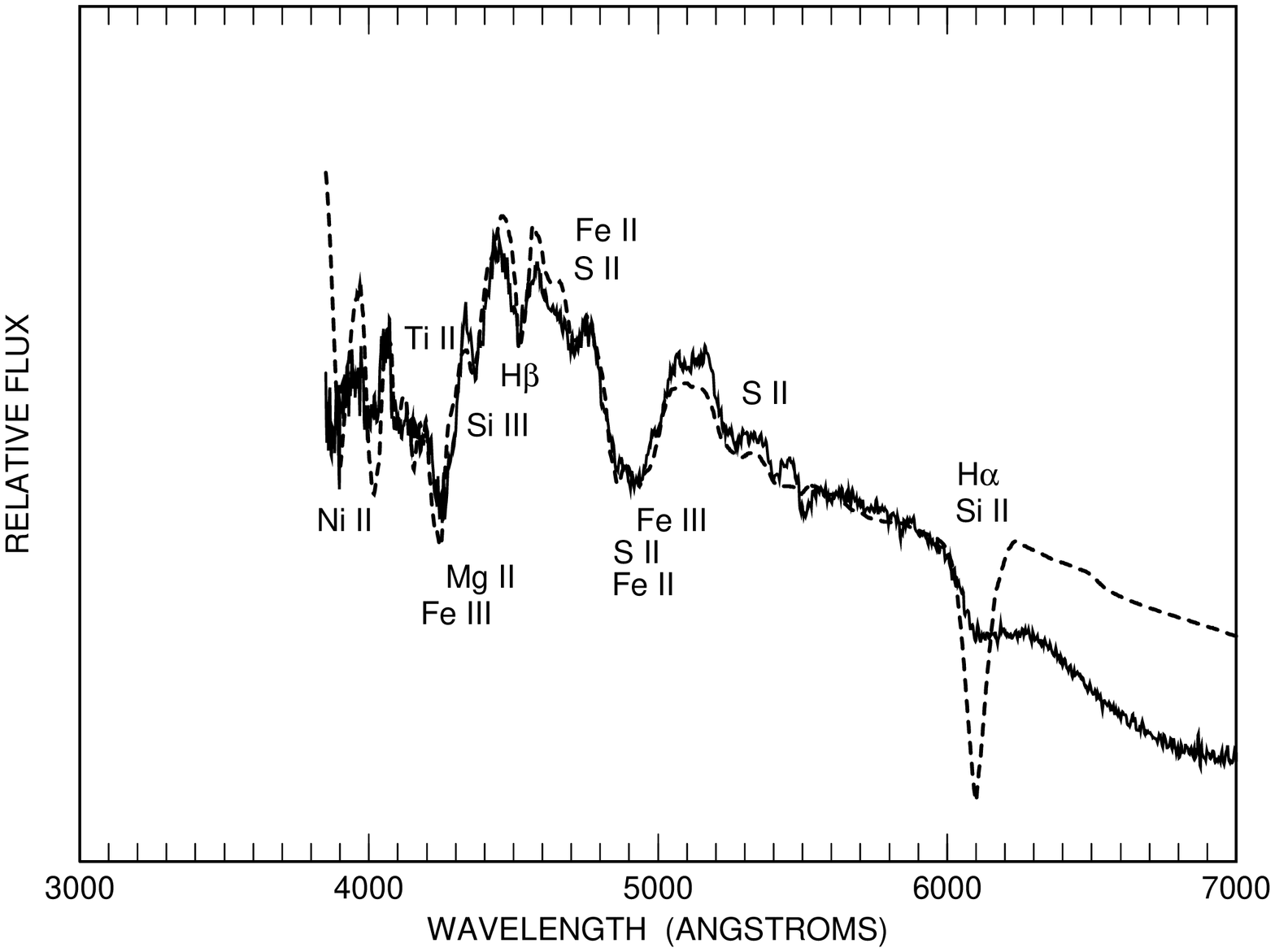}
\caption{The day~$-3$ spectrum of SN~2000cx ({\sl solid line}) is
  compared with a synthetic spectrum ({\sl dashed line})
  that has $v_{phot}=12,000$ \kms, $T_{bb}=9000$~K, and contains lines
  of nine ions.}
\end{figure}

\clearpage

\begin{figure}
\includegraphics[width=.8\textwidth,angle=270]{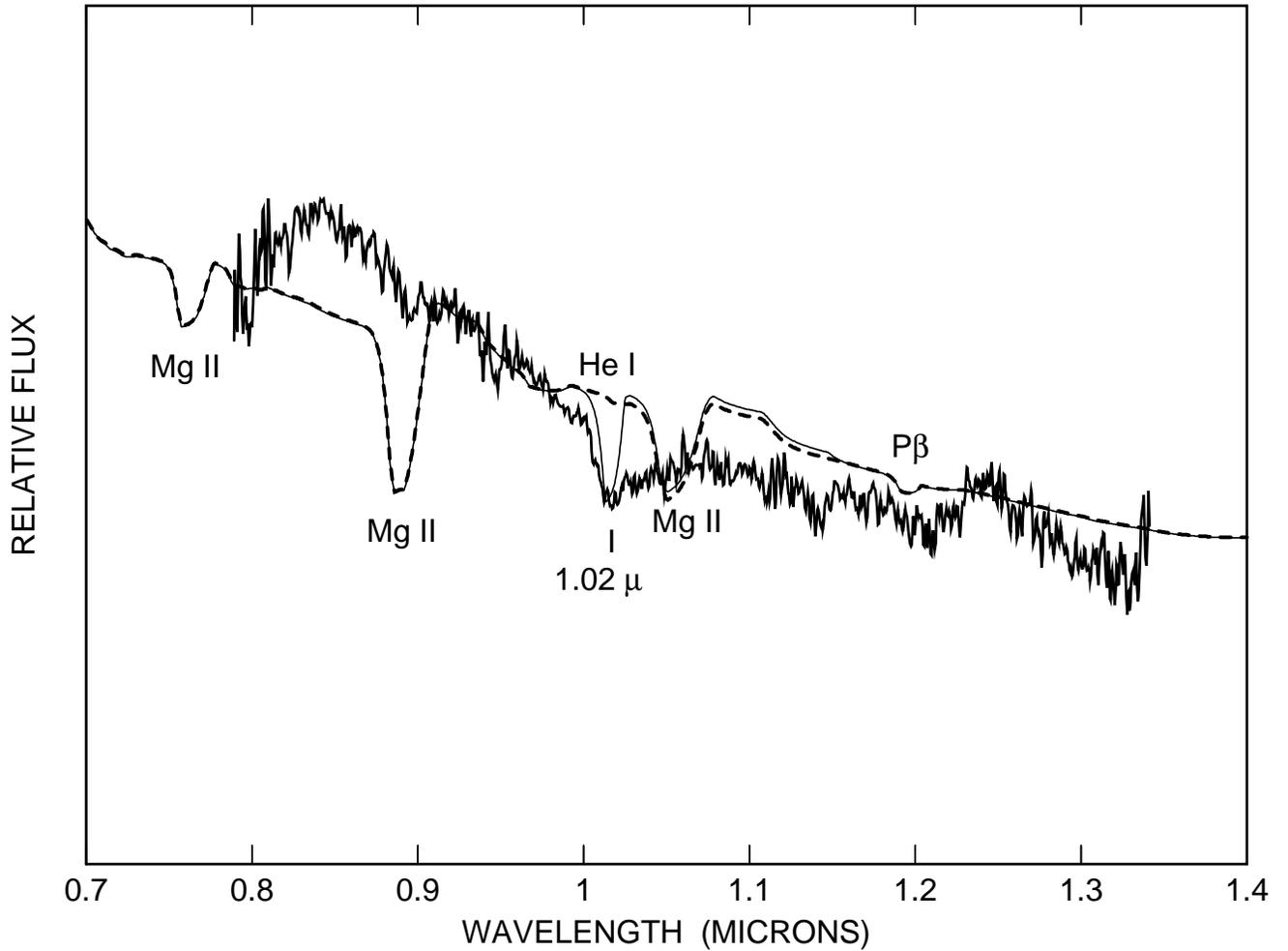}
\caption{The day~$-5.5$ short--wavelength infrared spectrum of
  SN~2000cx ({\sl solid line}) is compared with an extension of the
  synthetic spectrum ({\sl dashed line}) of Figure~15 for the day~$-3$
  optical spectrum, and the same synthetic spectrum but with the
  inclusion of He~I lines detached at 22,000 \kms\ ({\sl thin solid
  line}).}
\end{figure}

\clearpage
 
\begin{deluxetable}{lcccccr}
\footnotesize
\tablecaption{Input Parameters for Figure 4 (Day 2) \label{table1}}
\tablewidth{0pt}
\tablehead{
\colhead{ion} &
\colhead{$\lambda$(ref)} &
\colhead{$\tau$(ref)} & 
\colhead{$v_{min}$} &
\colhead{$v_{max}$} &
\colhead{$v_e$} &
\colhead{$T_{exc}$} \\
\colhead{} &
\colhead{(\AA)} &
\colhead{} &
\colhead{(\kms)} &
\colhead{(\kms)} &
\colhead{(\kms)} &
\colhead{(K)} 
} 
\startdata
Mg II  &$\lambda$4481 &1.8  & 13000 &$\infty$ &1000    & 7000 \\
Si II  &$\lambda$6347 &1.5  & 13000 &$\infty$ &1000    & 7000 \\
S II   &$\lambda$5454 &1.3  & 13000 &$\infty$ &1000    &10000 \\
Ca II  &$\lambda$3934 &2.0  & 11000 & 23000   &8000    & 5000 \\
Ca II  &$\lambda$3934 &0.7  & 23000 &$\infty$ &3000    & 5000 \\
Ti II  &$\lambda$4550 &0.4  & 23000 &$\infty$ &1000    & 5000 \\
Fe II  &$\lambda$5018 &0.25 & 21000 &$\infty$ &1000    & 5000 \\
Fe III &$\lambda$5156 &2.0  & 13000 &$\infty$ &1000    &14000 \\
Co II  &$\lambda$4161 &1.5  & 13000 &$\infty$ &1000    & 7000 \\
Ni II  &$\lambda$4067 &3.0  & 13000 &$\infty$ &1000    & 7000 \\

\enddata
\end{deluxetable}
\clearpage

\begin{deluxetable}{lcccccr}
\footnotesize
\tablecaption{Input Parameters for Figure 11 (Day 7) \label{table2}}
\tablewidth{0pt}
\tablehead{
\colhead{ion} &
\colhead{$\lambda$(ref)} &
\colhead{$\tau$(ref)} & 
\colhead{$v_{min}$} &
\colhead{$v_{max}$} &
\colhead{$v_e$} &
\colhead{$T_{exc}$} \\
\colhead{} &
\colhead{(\AA)} &
\colhead{} &
\colhead{(\kms)} &
\colhead{(\kms)} &
\colhead{(\kms)} &
\colhead{(K)}
} 
\startdata
H I    &$\lambda$6563 &7.0   & 22000 & $\infty$ & 1000    & 5000 \\
Mg II  &$\lambda$4481 &2.5   & 13000 & $\infty$ & 1000    & 7000 \\
Si II  &$\lambda$6347 &1.0   & 13000 & $\infty$ & 1000    & 8000 \\
S II   &$\lambda$5454 &1.0   & 13000 & $\infty$ & 1000    &10000 \\
Ca II  &$\lambda$3934 &2.5   & 11000 & 23000    & 8000    & 5000 \\
Ca II  &$\lambda$3934 &0.8   & 23000 & $\infty$ & 3000    & 5000 \\
Ti II  &$\lambda$4550 &0.6   & 23000 & $\infty$ & 1000    & 5000 \\
Fe II  &$\lambda$5018 &0.6   & 19000 & $\infty$ & 2000    & 7000 \\
Fe III &$\lambda$5156 &2.5   & 13000 & $\infty$ & 1000    &14000 \\
Co II  &$\lambda$4161 &2.0   & 13000 & $\infty$ & 1000    & 7000 \\
Ni II  &$\lambda$4067 &4.0   & 13000 & $\infty$ & 1000    & 7000 \\

\enddata
\end{deluxetable}
\clearpage

\begin{deluxetable}{lcccccr}
\footnotesize
\tablecaption{Input Parameters for Figure 14 (Day 15) \label{table3}}
\tablewidth{0pt}
\tablehead{
\colhead{ion} &
\colhead{$\lambda$(ref)} &
\colhead{$\tau$(ref)} & 
\colhead{$v_{min}$} &
\colhead{$v_{max}$} &
\colhead{$v_e$} &
\colhead{$T_{exc}$} \\
\colhead{} &
\colhead{(\AA)} &
\colhead{} &
\colhead{(\kms)} &
\colhead{(\kms)} &
\colhead{(\kms)} &
\colhead{(K)}
} 
\startdata
H I    &$\lambda$6563 &4.0  & 22000 & $\infty$ & 1000    & 5000 \\
Na I   &$\lambda$5892 &1.5  & 10000 & $\infty$ & 7000    & 5000 \\
Si II  &$\lambda$6347 &2.0  & 13000 & $\infty$ & 1000    & 7000 \\
Ti II  &$\lambda$4550 &0.3  & 23000 & $\infty$ & 1000    & 5000 \\
Fe II  &$\lambda$5018 &0.7  & 11000 & 23000    & 20000   & 7000 \\
Fe III &$\lambda$5156 &2.5  & 10000 & $\infty$ & 1000    & 14000 \\
Co II  &$\lambda$4161 &1.0  & 13000 & $\infty$ & 1000    & 7000 \\

\enddata
\end{deluxetable}
\clearpage

\begin{deluxetable}{lcccccr}
\footnotesize
\tablecaption{Input Parameters for Figure 17 (Day 32) \label{table4}}
\tablewidth{0pt}
\tablehead{
\colhead{ion} &
\colhead{$\lambda$(ref)} &
\colhead{$\tau$(ref)} & 
\colhead{$v_{min}$} &
\colhead{$v_{max}$} &
\colhead{$v_e$} &
\colhead{$T_{exc}$} \\
\colhead{} &
\colhead{(\AA)} &
\colhead{} &
\colhead{(\kms)} &
\colhead{(\kms)} &
\colhead{(\kms)} &
\colhead{(K)}
} 
\startdata
Na I   &$\lambda$5892 &1.5   & 12000 & $\infty$ & 1000    & 5000 \\
Si II  &$\lambda$6347 &2.0   & 13000 & $\infty$ & 1000    & 7000 \\
Ca II  &$\lambda$3934 &4.0   & 11000 & 20000    & 3000    & 5000 \\
Ca II  &$\lambda$3934 &2.0   & 20000 & $\infty$ & 1000    & 5000 \\
Fe II  &$\lambda$5018 &100   & 11000 & 13000    & 2000    & 10000 \\
Fe II  &$\lambda$5018 &2.0   & 13000 & $\infty$ & 1000    & 7000 \\

\enddata
\end{deluxetable}
\clearpage

\begin{deluxetable}{lcccccr}
\footnotesize
\tablecaption{Input Parameters for Figure 20 (Day $-3$) \label{table5}}
\tablewidth{0pt}
\tablehead{
\colhead{ion} &
\colhead{$\lambda$(ref)} &
\colhead{$\tau$(ref)} & 
\colhead{$v_{min}$} &
\colhead{$v_{max}$} &
\colhead{$v_e$} &
\colhead{$T_{exc}$} \\
\colhead{} &
\colhead{(\AA)} &
\colhead{} &
\colhead{(\kms)} &
\colhead{(\kms)} &
\colhead{(\kms)} &
\colhead{(K)}
} 
\startdata
H I    &$\lambda$6563 &7.0   & 23000 & $\infty$ & 1000    &  5000 \\
Mg II  &$\lambda$4481 &1.0   & 13000 & $\infty$ & 1000    &  7000 \\
Si II  &$\lambda$6347 &0.2   & 14000 & $\infty$ & 1000    &  7000 \\
Si III &$\lambda$4550 &1.5   & 14000 & $\infty$ & 1000    & 14000 \\
S II   &$\lambda$5454 &0.2   & 14000 & $\infty$ & 1000    & 10000 \\
Ti II  &$\lambda$4550 &0.7   & 23000 & $\infty$ & 1000    &  5000 \\
Fe II  &$\lambda$5018 &0.5   & 21000 & $\infty$ & 1000    &  7000 \\
Fe III &$\lambda$5156 &1.0   & 14000 & $\infty$ & 1000    & 14000 \\
Ni II  &$\lambda$4067 &1.0   & 13000 & $\infty$ & 1000    &  7000 \\

\enddata
\end{deluxetable}
\clearpage

\end{document}